\documentclass[aps,prb,reprint,twocolumn]{revtex4-2}

\usepackage{graphicx}
\usepackage{bm}
\usepackage{hyperref}
\usepackage{amsmath}
\usepackage{amssymb}
\usepackage{siunitx}

\usepackage{comment}
\usepackage[table]{xcolor}
\usepackage{array}
\usepackage{multirow}
\usepackage{physics}
\usepackage{url}

\begin{document}
\title{Parent Berry curvature and the ideal anomalous Hall crystal}
\date{\today}
\author{Tixuan Tan}
\affiliation{Department of Physics, Stanford University, Stanford, CA 94305, USA}
\author{Trithep Devakul}
\email{tdevakul@stanford.edu}  
\affiliation{Department of Physics, Stanford University, Stanford, CA 94305, USA}
\begin{abstract}
We study a model of electrons moving in a parent band of uniform Berry curvature.
At sufficiently high parent Berry curvature, we show that strong repulsive interactions generically lead to the formation of an anomalous Hall crystal: a topological state with spontaneously broken continuous translation symmetry.
Our results are established via a mapping to a problem of Wigner crystallization in a regular 2D electron gas.
Interestingly, we find that a periodic electrostatic potential induces a competing state with opposite Chern number.
Our theory offers a unified perspective for understanding several aspects of the recently observed integer and fractional quantum anomalous Hall effects in rhombohedral multilayer graphene and provides a recipe for engineering new topological states.
\end{abstract}
\maketitle

\section{Introduction}

Electron interactions and topology can cooperate to create fascinating states of matter.
This has been made abundantly clear in the field of 2D materials, where much exciting progress has been made in the study of topological flat band systems~\cite{andrei2021marvels,mak2022semiconductor}.
In materials like twisted bilayer graphene (TBG)~\cite{bistritzer2011moire,cao2018unconventional,cao2018correlated} or twisted transition metal dichalcogenides (TMD) ~\cite{wu2019topological,devakul2021magic}, a moir\'e superlattice leads to the formation of topological flat minibands which, when partially occupied, provides an ideal setting for interactions and topology to dominate.
This has led to the recent experimental realization of exotic correlated topological states such as fractional Chern insulators ~\cite{liu2022recent,parameswaran2013fractional,bergholtz2013topological,tang2011high,neupert2011fractional,regnault2011fractional,sheng2011fractional} at zero magnetic field ~\cite{park2023observation,xu2023observation,cai2023signatures,zeng2023thermodynamic,lu2024fractional}.
Very recent observations of the fractional quantum spin Hall effect ~\cite{kang2024observation} is further evidence that novel states of matter, never seen before, are now becoming reality in topological flat band systems.

\begin{figure}[t]
\begin{center}
\includegraphics[width=0.48\textwidth]{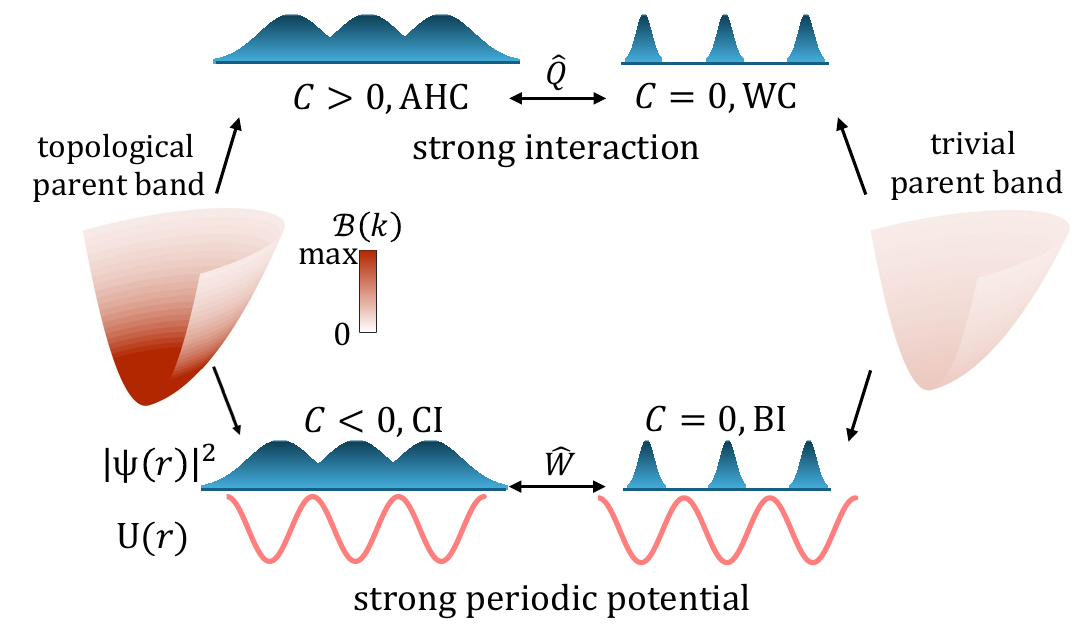}
\end{center}
\caption{Summary of results and methodology. 
A topological parent band with a high concentration of Berry curvature $\mathcal{B}>0$
is driven into a $C>0$ anomalous Hall crystal (AHC) by strong interactions, or a $C<0$ Chern insulator (CI) by a periodic potential.
These results are obtained by utilizing a unitary mapping to a model with a trivial parent band. The trivial parent band is driven to a Wigner crystal (WC) by interactions, or a band insulator (BI) by a periodic potential.
The AHC is mapped to the WC by a unitary transformation $\hat{Q}$, while the CI is mapped to the BI by a different unitary $\hat{W}$.
}
\label{fig:schematic}
\end{figure}

There is another setting, without flat bands (at the single-particle level), in which electronic topology and interactions can be at the forefront.  
Consider a system of electrons at a semiconductor band edge, which we will refer to as the ``parent'' band.
Here, the density of electrons is low enough that the atomic Brillouin zone is irrelevant.
When the parent band is topologically trivial, the system is a usual 2D electron gas (2DEG).  
In the presence of strong repulsive interactions, the 2DEG can spontaneously crystallize to form a Wigner crystal state ~\cite{wigner1934interaction}.
However, as will be made clear, topology becomes unavoidable when the parent band carries a high concentration of Berry curvature near its band edge, as illustrated in Fig~\ref{fig:schematic}.  
In the presence of strong repulsive interactions, it is possible that electrons in this topological parent band may spontaneously crystallize to form an exotic ``anomalous Hall crystal'' (AHC), a topological version of the Wigner crystal with an integer Chern number ~\cite{kivelson1986cooperative,Halperin1986compatibility,kivelson1987cooperative,tevsanovic1989hall} at zero magnetic field, arising from a synergistic interplay of electronic interaction and parent band topology (which can also be viewed as a supersolid~\cite{Ceperley2004Ring,Feynman1953transition,Leggett1970Can}).
{ We emphasize that our focus is on states with spontaneously broken continuous translation symmetry, rather than those with broken discrete lattice symmetries~\cite{polshyn2022topological}. }

One class of materials with highly concentrated Berry curvature is rhombohedral multilayer graphene~\cite{chen2020tunable,Zhang2019nearly,Correlated2023Han,lu2024fractional,han2023large,Zhou_2021super,Zhou_2021half,de_la_Barrera_2022,Seiler_2022,Zhou_2022isospin,liu2023spontaneous,han2023orbital}.
We are motivated by the recent experimental observation of the integer and fractional quantum anomalous Hall (QAH) effects in rhombohedral pentalayer graphene (R5G) aligned with hBN~\cite{lu2024fractional}, and theoretical works following~\cite{dong2023theory,zhou2023fractional,dong2023anomalous,guo2023theory,kwan2023moir} (see ~\cite{Vishwanath2023zerocomment,Parameswaran2024comment} for a summary).
The single-particle miniband structure of R5G/hBN does not possess an isolated flat band~\cite{Park2023topologicalflat,jung2015origin,Jungjeil2017bandmodel,Wallband2013generic},
suggesting that the underlying mechanism for fractionalization is unconventional (distinct from twisted MoTe$_2$~\cite{park2023observation,xu2023observation,cai2023signatures,zeng2023thermodynamic}, which can essentially be understood as quantum Hall physics in a flat Chern band).
% This suggests that the underlying mechanism for the QAH effects in R5G/hBN is entirely different from that of twisted MoTe$_2$~\cite{park2023observation,xu2023observation,cai2023signatures,zeng2023thermodynamic} (which can essentially be understood as quantum Hall physics in a topological flat band).
Theoretical studies find that these topological state persists even in the absence of a moir\'e superlattice~\cite{zhou2023fractional,dong2023anomalous,kwan2023moir}, suggesting that the underlying mechanism is related to anomalous Hall crystallization. % and that the parent band perspective is the natural starting point.
However, the complexity of the microscopic model and subtleties in the treatment of electronic interactions~\cite{kwan2023moir} pose a serious obstacle to developing a general theoretical understanding of the origin of this state.  
Like the chiral limit ~\cite{Tarnopolsky2019origin} or the heavy fermion picture ~\cite{Song2022Heavy} of TBG, there is a need for simple, controlled, and analytically tractable models that can capture the essence of this system and reveal the universal physics beneath.

% {\color{green}
% In fact, robust correlated topological states are only observed when the electrons are localized on the layer \emph{furthest} from the BN-induced moir\'e, implying that a strong moir\'e potential might actually be deleterious.  
% Intriguingly, Hartree-Fock (HF) theory finds the integer QAH state even without the moir\'e~\cite{zhou2023fractional,dong2023anomalous,kwan2023moir}, implying that the state may be adiabatically connected to an AHC.
% This suggests that the parent band perspective, without moir\'e, is the natural starting point. 
% Not only this, numerical diagonalization shows that partially filling the HF bands lead to fractional Chern insulators, suggesting that its quantum geometry~\cite{resta2011insulating} is sufficiently ``ideal'' for the realization of such states~\cite{parameswaran2013fractional,roy2014band,Wang2021exact,ledwith2022vortexability,Claassen2015position,Ledwith2020fractionalchern,Lee2017engineering,jackson2015geometric,BrunoKahler2021}.
% However, the complexity of the microscopic model and subtleties in the treatment of electronic interactions~\cite{kwan2023moir} pose a serious obstacle to developing a general theoretical understanding of the origin of this state.  
% Like the chiral limit ~\cite{Tarnopolsky2019origin} or the heavy fermion picture ~\cite{Song2022Heavy} of TBG, there is a need for simple, controlled, and analytically tractable models that can capture the essence of this system and reveal the universal physics beneath.
% }

Our theory fulfills this role.
We explain these puzzling observations as natural consequences of strong electronic interactions in the presence of high parent Berry curvature.

We first introduce and study a model of an idealized parent band with a uniform, continuously tunable, parent Berry curvature. 
This model allows us to isolate the effect of parent Berry curvature from all other extraneous details.
% We propose that this idealized model captures the essential aspects of the AHC.
In this model, a strong repulsive interaction and/or periodic scalar potential leads to gapped states with interesting properties and enough structure that exact statements can be made about their topology in certain limits.

We show that in the presence of a high parent Berry curvature $\mathcal{B}$, both strong interactions or a periodic potential generically drive the system into a topological state with non-zero Chern number.  
Consider a system of spinless electrons with uniform parent Berry curvature $\mathcal{B}\approx 2\pi N/\Omega_{\mathrm{BZ}}$, where $N$ is a positive integer and $\Omega_{\mathrm{BZ}}=4\pi^2 n$ is the area of a Brillouin zone (chosen such that the electron density $n$ corresponds to unit filling).
We demonstrate that repulsive interactions will drive the system into an AHC with Chern number $C=N$, while a commensurate periodic potential will instead lead to a Chern insulator with $C=-N$.
Thus, there is very different behavior depending on whether the gap is interaction or potential-induced!  
Our results reveal that the two effects, repulsive interactions and a periodic potential, generically compete and drive the system to two fixed points with different topology.

These statements are established by utilizing a unitary mapping of the second-quantized Hamiltonian of a topological parent band model with $\mathcal{B}=2\pi N/\Omega_{\mathrm{BZ}}$ to a representative model with a trivial parent band $\mathcal{B}=0$.
The general methodology is illustrated in Fig~\ref{fig:schematic}.
The mapping makes use of the $2\pi$ flux periodicity of an effective momentum space Hofstadter model.  
Since the $\mathcal{B}=0$ state forms a trivial Wigner crystal state in the strongly interacting limit, or a trivial band insulator in a periodic potential, the properties of the $\mathcal{B}=2\pi N/\Omega_{\mathrm{BZ}}$ can be deduced.
The mapping transforms the Chern number in a simple way, giving rise to the $C=\pm N$ states.

Surprisingly, our results imply a direct relation between a Wigner crystal, which is adiabatically connected to a classical state of point-like electrons, 
and the AHC, which is a quantum strongly interacting topological state.  
We exploit this mapping to write down an explicit wavefunction ansatz for the AHC state that is accurate in the limit of strong repulsive interactions.
In this limit, we show that the quantum geometry~\cite{resta2011insulating} of the quasiparticle bands of the AHC becomes perfectly ``ideal''~\cite{parameswaran2013fractional,roy2014band,wang2021exact,ledwith2022vortexability,claassen2015position,Ledwith2020fractionalchern,Lee2017engineering,jackson2015geometric,BrunoKahler2021} for the realization of fractionalized phases.

Finally, although we study an idealized model, our results provide a powerful guiding perspective for understanding real systems.
When the parent band features a concentrated Berry curvature near the band edge, our analysis should apply as long as the relevant low-energy states involved are within the region of high Berry curvature.
{ We demonstrate this explicitly for rhombohedral multilayer graphene. }
We expect that our model, and the methodology developed in this work, will be useful for future studies into the role of Berry curvature in many-body physics.

\section{Model}\label{sec:model}

\subsection{General Hamiltonian}
We consider general Hamiltonian describing electrons projected to a single parent band, potentially in the presence of a periodic electrostatic potential $U(\bm{r})$ and density-density interactions $V(\bm{r})$,
\begin{equation}
\hat{\mathcal{H}}=\hat{\mathcal{H}_0} + \hat{\mathcal{H}}_{\mathrm{pot}} + \hat{\mathcal{H}}_{\mathrm{int}}
\label{eq:fullham}
\end{equation}
In second quantized notation, the kinetic term is
\begin{equation}
\hat{\mathcal{H}_0}=\sum_{\bm{k}} {c}_{\bm{k}}^\dagger \mathcal{E}(\bm{k}) {c}_{\bm{k}}
\end{equation}
where $c_{\bm{k}}^{\dagger}$ creates an electron at momentum $\bm{k}$ in the parent band, {denoted by $\ket{\bm{k}}$}, and $\mathcal{E}(\bm{k})$ is the parent band dispersion.
The effect of a periodic potential $U(\bm{r})=\sum_{\bm{g}}U_{\bm{g}}e^{i\bm{g}\cdot\bm{r}}$ gives rise to the term
\begin{equation}
\hat{\mathcal{H}}_{\mathrm{pot}}=\sum_{\bm{k},\bm{k}^\prime} {c}_{\bm{k}^\prime}^\dagger U_{\bm{k}^\prime-\bm{k}}  \mathcal{F}(\bm{k}^\prime,\bm{k}) {c}_{\bm{k}}
\end{equation}
% where $\mathcal{F}(\bm{k^\prime},\bm{k})=\braket{u_{\bm{k}^\prime}}{u_{\bm{k}}}$ is the form factor, and $\ket{u_{\bm{k}}}$ are the periodic parts of the parent band Bloch wavefunctions.
{
where $\mathcal{F}(\bm{k^\prime},\bm{k})=\bra{\bm{k}^\prime}e^{i(\bm{k}^\prime-\bm{k})\cdot\bm{r}}\ket{\bm{k}}$ is the form factor. }
We consider a density-density interaction 
\begin{equation}
\hat{\mathcal{H}}_{\mathrm{int}}=\frac{1}{2A}\sum_{\bm{k}_1\bm{k}_2\bm{k}_3\bm{k}_4} \tilde{V}_{\bm{k}_1\bm{k}_2\bm{k}_3\bm{k}_4}
{c}_{\bm{k}_1}^\dagger {c}_{\bm{k}_2}^\dagger  {c}_{\bm{k}_3} c_{\bm{k}_4}\label{eq:Hint}
\end{equation}
where $A$ is the total area of the system, and
\begin{equation}
\begin{split}
{V}_{\bm{k}_1\bm{k}_2\bm{k}_3\bm{k}_4} =& V(\bm{k}_1-\bm{k}_4)
\mathcal{F}(\bm{k}_1,\bm{k}_4)
\mathcal{F}(\bm{k}_2,\bm{k}_3)\\
& \times \delta_{\bm{k}_1+\bm{k}_2-\bm{k}_3-\bm{k}_4}
\end{split}
\end{equation}
where $V(\bm{q})=\int  V(\bm{r}) e^{-i\bm{q}\cdot\bm{r}}d\bm{r}$ is the Fourier transform of the real-space density-density interaction.
This model thus far is entirely general.
The quantum geometry of the parent band are encoded in the form factors $\mathcal{F}$.
The remainder of this section will discuss a particular microscopic model giving rise to the parent band.

{
\subsection{Ideal parent band}
 We study a model of the simplest possible parent band with uniform Berry curvature $\mathcal{B}$, which is taken to be a model parameter.
We first state the essential properties of this model that enter into the second quantized Hamiltonian.
We define the operators $c_{\bm{k}}^{\dagger[\mathcal{B}]}$ to create a fermion in the state $\ket{{\bm{k}}}=e^{i\bm{k}\cdot\bm{r}}\ket{s_{\bm{k}}^{\mathcal{B}}}$ in the ideal parent band, where $\ket{s_{\bm{k}}^{\mathcal{B}}}$ is an internal spinor satisfying $\braket{s_{\bm{k}^\prime}^{\mathcal{B}}}{s_{\bm{k}}^{\mathcal{B}}}=\mathcal{F}_{\mathcal{B}}(\bm{k}^\prime,\bm{k})$, with the form factor
\begin{equation}
\mathcal{F}_{\mathcal{B}}(\bm{k}^\prime,\bm{k})=\exp{-\frac{\mathcal{B}}{4}(|\bm{k}^\prime-\bm{k}|^2+2i\bm{k}^\prime\cross\bm{k})}
\label{eq:Funiform}
\end{equation}
where $\bm{k}\cross\bm{q}\equiv k_x q_y - k_y q_x$.
We take the single-particle kinetic energy to be simply $\mathcal{E}(\bm{k})=\frac{|\bm{k}|^2}{2m}$, although the precise dispersion is unimportant.
We will typically neglect the $\mathcal{B}$ superscript on $c^{\dagger}_{\bm{k}}$ when it is obvious.
% The form factor given in Eq~\ref{eq:Funiform} fully defines the second quantized Hamiltonian that will be studied in this paper.

The form factor consists of a geometric factor $e^{-\frac{\mathcal{B}}{4}|\bm{k}^\prime-\bm{k}|^2}$ arising due to the ``quantum distance'' between the two states, and a phase factor which can be interpreted as a momentum space Aharonov-Bohm factor $\arg \mathcal{F}_{\mathcal{B}}(\bm{k}^\prime,\bm{k})=\int_{\bm{k}}^{\bm{k}^\prime}\bm{A}(\bm{q})\cdot d\bm{q}$ with the Berry connection $\bm{A}(\bm{k})=\frac{\mathcal{B}}{2}(-k_y,k_x)$ playing the role of the vector potential in the symmetric gauge.

It can be verified that Eq~\ref{eq:Funiform} leads to the desired uniform Berry curvature
\begin{equation}
\begin{split}
\mathcal{B}(\bm{k})&\equiv -\Im\sum_{\mu,\nu}[\epsilon_{\mu\nu}\braket{\partial_{k_\mu}s_{\bm{k}}^{\mathcal{B}}}{\partial_{k_\nu}s_{\bm{k}}^{\mathcal{B}}}]
% =\frac{4SM^2}{(|\bm{k}|^2+M^2)^2}
% &=-\Im\sum_{\mu,\nu} \left.\epsilon_{\mu\nu} \partial_{k_\mu^\prime}\partial_{k_{\nu}}\mathcal{F}_{\mathcal{B}}(\bm{k}^\prime,\bm{k})\right|_{\bm{k^\prime=\bm{k}}}=\mathcal{B}
=\mathcal{B}
\end{split}
\label{eq:UnifBerryDistrib}
\end{equation}
where $\mu,\nu\in\{x,y\}$ and $\epsilon_{xy}=-\epsilon_{yx}=1$.
% Not only this, numerical diagonalization shows that partially filling the HF bands lead to fractional Chern insulators, suggesting that its quantum geometry~\cite{resta2011insulating} is sufficiently ``ideal'' for the realization of such states~\cite{parameswaran2013fractional,roy2014band,Wang2021exact,ledwith2022vortexability,Claassen2015position,Ledwith2020fractionalchern,Lee2017engineering,jackson2015geometric,BrunoKahler2021}.
To gain insight into the quantum geometry of this band, we may also compute the Fubini-Study metric
\begin{align}
g^{\mathrm{FS}}_{\mu\nu}(\bm{k})&=\Re\left[\braket{\partial_{k_\mu} s_{\bm{k}}^{\mathcal{B}}}{\partial_{k_\nu} s_{\bm{k}}^{\mathcal{B}}} - 
\braket{\partial_{k_\mu} s_{\bm{k}}^{\mathcal{B}}}{s_{\bm{k}}^{\mathcal{B}}}
\braket{s_{\bm{k}}^{\mathcal{B}}}{\partial_{k_\nu} s_{\bm{k}}^{\mathcal{B}}}
\right]\nonumber\\
& = \frac{1}{2}\mathcal{B}\delta_{\mu\nu}
\end{align}
which exactly saturates both the
trace $\Tr[g^{\mathrm{FS}}_{\mu\nu}(\bm{k})]\geq |\mathcal{B}(\bm{k})|$ and determinant $\det[g^{\mathrm{FS}}_{\mu\nu}(\bm{k})]\geq\frac{1}{4}|\mathcal{B}(\bm{k})|^2$ bounds~\cite{parameswaran2013fractional,roy2014band,jackson2015geometric,claassen2015position}.
In this sense, the quantum metric is the ``minimal'' given the Berry curvature distribution.
% Flat Chern bands satisfying the trace condition have been called ``ideal'' for the realization of fractionalized phases~\cite{parameswaran2013fractional,roy2014band,Wang2021exact,ledwith2022vortexability,Claassen2015position,Ledwith2020fractionalchern,Lee2017engineering,jackson2015geometric,BrunoKahler2021}.
This model therefore allows for the systematic study of the effect of parent Berry curvature, with minimal contributions from other quantum geometric effects.

We note that Eq~\ref{eq:Funiform} is identical to that obtained for the lowest Landau level with magnetic length $\ell_{B}^2=\mathcal{B}$~\cite{supplement}.
This parent band can therefore be thought of as a dispersive ``big'' Landau level in which momentum is unbounded, or equivalently, as an ``infinite'' ideal Chern band.
In the same way that a Landau level is the simplest Chern band, we argue that this model is the simplest parent band understanding the AHC.

\subsection{Microscopic realizations}\label{sec:microscopic}
The remainder of this section will discuss microscopic realizations of the ideal parent band, namely, a concrete form for the spinors $\ket{s_{\bm{k}}^{\mathcal{B}}}$.  
These details will not be necessary to derive any of the results of this paper, which rely only on $\mathcal{F}_{\mathcal{B}}$ in Eq~\ref{eq:Funiform}, so the reader may wish to proceed directly to Sec~\ref{sec:periodic}.

We present two microscopic approaches. 
First, we will present a model in which electrons carry a $2S$-dimensional internal degree of freedom, which gives rise to a tunable Berry curvature distribution; $\mathcal{F}_{\mathcal{B}}$ is realized asymptotically in the limit of uniform Berry curvature and large $S$.
Next, we will present a model in which the fermions carry an unbounded internal degree of freedom, which gives rise to $\mathcal{F}_{\mathcal{B}}$ directly.

We emphasize again that the details of these microscopic models are not used in any way in the remainder of this paper. The purpose of presenting these models is so that they may be useful in future studies.

\subsubsection{Multifold band inversion model}\label{sec:multifold}
We introduce the ``multifold band inversion'' model,
% \begin{equation}
% \hat{\mathcal{H}}^{\mathrm{MBI}} = \sum_{\bm{k}}{a}^\dagger_{\bm{k}\sigma^\prime} H^{\mathrm{MBI}}_{\sigma^\prime\sigma}(\bm{k}){a}_{\bm{k}\sigma}
% \end{equation}
% where $a^\dagger_{\bm{k}\sigma}$ creates a fermion at momentum $k$ with an internal degree of freedom $\sigma=-S,\dots,S$ (thus describing an effective spin-$S$ fermion, although the microscopic origin is unimportant).
described by the Bloch Hamiltonian
\begin{equation}
H^{\mathrm{MBI}}_{\sigma^\prime\sigma}(\bm{k}) = h_0(\bm{k})\delta_{\sigma^\prime\sigma}+[\bm{h}(\bm{k})\cdot\bm{S}]_{\sigma^\prime\sigma}
\label{eq:MBIHam}
\end{equation}
where $\sigma=-S,\dots,S$ with half-integer $S$, 
\begin{equation}
\bm{h}(\bm{k}) = \left(vk_x,vk_y,\frac{|\bm{k}|^2}{m}+\delta\right)
\end{equation}
and $h_0(\bm{k})=(\frac{1}{2}-S)|\bm{h}(\bm{k})|$. 
Here, $\bm{S}=(S^x,S^y,S^z)$ are the standard spin-$S$ matrices 
satisfying $[S^x,S^y]=iS^z$ and cyclic permutations thereof, and $\bm{S}^2=S(S+1)$.
For $S=\frac{1}{2}$, $H^{\mathrm{MBI}}(\bm{k})$ is the two-band low-energy model for a topological band inversion ~\cite{BHZ}.

At each momentum $\bm{k}$, $H^{\mathrm{MBI}}(\bm{k})$ simply acts as a spin Zeeman field $\bm{h}(\bm{k})$, thus the eigenvalues are given by 
\begin{equation}
E^{\mathrm{MBI}}_n(\bm{k})=h_0(\bm{k})+(S-n)|\bm{h}(\bm{k})|=\left(\frac{1}{2}-n\right)|\bm{h}(\bm{k})|.
\end{equation}
where $n=0,\dots,2S$ is the band index.  
We shall focus on the highest energy band as our parent band, $E^{\mathrm{MBI}}_0(\bm{k})$. 
The purpose of the $h_0(\bm{k})$ term is to ensure that all other bands have negative energy, and can be assumed to be fully filled and inert.
The $n=0$ band is topologically trivial for $\delta>0$, and becomes topological for $\delta<0$ with a Chern number of $C=2S$.
The point $\delta=0$ describes a multifold fermion critical point in which all $2S+1$ bands touch at $\bm{k}=0$.

% \subsubsection{Ideal point and large $S$ limit}
We now consider a special ``ideal'' point by setting $\delta=\delta^*\equiv-{mv^2/4}$.
At this point, $H^{\mathrm{MBI}}$ has several desirable properties.
The energy of the highest band becomes parabolic, $E_0(\bm{k})=\frac{k^2}{2m}+\frac{m v^2}{8}$.
The corresponding eigenvector of $H^{\mathrm{MBI}}(\bm{k})$ in the eigenbasis of $S^z$ is
\begin{equation}
    v_{\sigma}(\bm{k})=\sqrt{\frac{(2S)!}{(S-\sigma)!(S+\sigma)!}}\frac{(k_x-ik_y)^{S+\sigma}M^{S-\sigma}}{(|\bm{k}|^2+M^2)^S}
    \label{eq:vsigma}
\end{equation}
where $M=mv/2$.
The eigenstates $\ket{s_{\bm{k}}}=\sum_{\sigma=-S}^{S}v_\sigma(\bm{k})\ket{\sigma}$ give rise to the form factors
\begin{equation}
\mathcal{F}(\bm{k}^\prime,\bm{k})=\frac{(M^2+\bm{k}^\prime\cdot\bm{k}-i\bm{k^\prime}\cross\bm{k})^{2S}}{(M^2+|\bm{k}^\prime|^2)^S(M^2+|\bm{k}|^2)^S}
\end{equation}
and the Berry curvature
\begin{equation}
\mathcal{B}(\bm{k})=\frac{4SM^2}{(|\bm{k}|^2+M^2)^2}
\label{eq:BerryDistrib}
\end{equation}
which integrates to a total Chern number of $C=\frac{1}{2\pi}\int \mathcal{B}(\bm{k})d\bm{k}=2S$ as advertised.
The Fubini-Study metric satisfies
$g^{\mathrm{FS}}_{\mu\nu}(\bm{k}) = \frac{1}{2}\mathcal{B}(\bm{k})\delta_{\mu\nu}$, thus satisfying the trace and determinant conditions at all $\bm{k}$.
% This model therefore allows for the systematic study of the effect of parent band Berry curvature, with minimal contributions from other quantum geometric effects.
The total magnitude of $\mathcal{B}(\bm{k})$ can be tuned by $S$, and the distribution can be tuned by $M$.  

We anticipate that this model will be useful for studying the effect of parent Berry curvature distribution, with minimal contributions from other effects.
This model allows for arbitrarily high total Berry flux, which can be made arbitrarily uniform or concentrated.
This should be contrasted with, say, a massive Dirac fermion in which the total Berry flux is limited to $\pi$.
More generally, in the effective two-band model of rhombohedral $N$-layer graphene in a displacement field~\cite{Min_2008}, $H(k)=\Delta \sigma^z + t(\hbar v/t)^N\left[(k_x-i k_y)^N \sigma^+ + h.c.\right]$, the total Berry flux is $N\pi$;
while the total Berry flux can be made arbitrarily high, the distribution becomes highly non-uniform (it takes a ring-like shape) and the quantum metric is not minimal.
Hence, $H^{\mathrm{MBI}}$ at the ideal point provides a controlled model for studying the effect of Berry curvature while minimizing other details.

In this work, we are interested in the case of a uniform positive Berry curvature distribution $\mathcal{B}(\bm{k})=\mathcal{B}>0$.
This can be achieved by setting $M=\sqrt{4S/\mathcal{B}}$ and then taking the limit $S\rightarrow\infty$ (keeping $m$ fixed).
In this limit, the form factors reduce to $\mathcal{F}_{\mathcal{B}}$ in Eq~\ref{eq:Funiform}.
% which consists of a geometric factor $e^{-\frac{\mathcal{B}}{4}|\bm{k}^\prime-\bm{k}|^2}$ arising due to the ``quantum distance'' between the two states, and a phase factor which can be interpreted as a momentum space Aharonov-Bohm factor $\arg \mathcal{F}_{\mathcal{B}}(\bm{k}^\prime,\bm{k})=\int_{\bm{k}}^{\bm{k}^\prime}\bm{A}(\bm{q})\cdot d\bm{q}$ with the Berry connection $\bm{A}(\bm{k})=\frac{\mathcal{B}}{2}(-k_y,k_x)$ playing the role of the vector potential in the symmetric gauge.
% We remark that Eq~\ref{eq:Funiform} is identical to that obtained for the lowest Landau level with magnetic length $\ell_{B}^2=\mathcal{B}$, but describes an unbounded spectrum of parabolically dispersing electrons.  

% We will utilize this limit as the parent band in the Hamiltonian Eq~\ref{eq:fullham}.  
% The single-particle dispersion in $\mathcal{H}_0$ is given by $\mathcal{E}(\bm{k})=\frac{|\bm{k}|^2}{2m}$ (a constant offset has been neglected), and the form factors in $\hat{\mathcal{H}}_{\mathrm{pot,int}}$ are given by Eq~\ref{eq:Funiform}.
% We define the operators $c_{\bm{k}}^{\dagger[\mathcal{B}]}$ to create a fermion in the microscopic state $e^{i\bm{k}\cdot\bm{r}}\ket{u_{\bm{k}}^{\mathcal{B}}}$ in the ideal parent band, which satisfy $\braket{u_{\bm{k}^\prime}^{\mathcal{B}}}{u_{\bm{k}}^{\mathcal{B}}}=\mathcal{F}_{\mathcal{B}}(\bm{k}^\prime,\bm{k})$.
% We will typically neglect the $\mathcal{B}$ superscript on $c^{\dagger}_{\bm{k}}$ when it is unimportant.
\subsubsection{Infinite Chern band model}
We now introduce an alternative microscopic realization of the ideal parent band that directly gives rise to a uniform Berry curvature.
We consider the ``infinite Chern band'', described by
% \begin{equation}
% \hat{\mathcal{H}}^{\mathrm{ICB}} = \sum_{\bm{k}}a_{\bm{k}n^\prime}H_{n^\prime n}^{\mathrm{ICB}}(\bm{k}) a_{\bm{k} n}
% \end{equation}
\begin{equation}
H^{\mathrm{ICB}}_{n^\prime n}(\bm{k}) = \frac{|\bm{k}|^2}{2m}\delta_{n^\prime n} + J L_{n^\prime n}(\bm{k})
\end{equation}
where $n=0,1,\dots,\infty$ is an unbounded internal degree of freedom.
Here, $L_{n^\prime n}(\bm{k}) =\bra{n^\prime}L(\bm{k})\ket{n}$, with
\begin{equation}
  L(\bm{k})=(b^\dagger -\sqrt{\mathcal{B}}k)(b-\sqrt{\mathcal{B}}k^*)
\end{equation}
where $k\equiv (k_x+ik_y)/\sqrt{2}$, and $b,b^\dagger$ are ladder operators acting on the basis states $\ket{n}$, satisfying the usual relations $[b,b^\dagger]=1$, $b\ket{n}=\sqrt{n}\ket{n-1}$,$b^\dagger\ket{n}=\sqrt{n+1}\ket{n+1}$.

To solve for the spectrum of $H^{\mathrm{ICB}}$, we notice that $L(\bm{k})=d^\dagger(\bm{k}) d(\bm{k})$, where $d(\bm{k})=b-\sqrt{\mathcal{B}}k^*$ is a shifted ladder operator.
The spectrum of $L$ therefore consists of the non-negative integers.  
The eigenvalues of $H^{\mathrm{ICB}}$ is simply $E_n(\bm{k})=\frac{|\bm{k}|^2}{2m}+nJ$. 
We take $J\rightarrow\infty$ to be large, and focus only on the lowest band $\mathcal{E}(\bm{k})=E_0(\bm{k})$ as our parent band.  
These eigenstates are annihilated by $d(\bm{k})$, so they are coherent states satisfying $b\ket{s_{\bm{k}}^{\mathcal{B}}}=\sqrt{\mathcal{B}}k^*\ket{s_{\bm{k}}^{\mathcal{B}}}$, given by
\begin{equation}
\ket{s_{\bm{k}}^{\mathcal{B}}}
=e^{\sqrt{\mathcal{B}}(k^*b^\dagger-kb)}\ket{0}
= e^{-\frac{\mathcal{B}}{4}|\bm{k}|^2} \sum_{n=0}^{\infty} \frac{(\sqrt{\mathcal{B}} k^*)^n}{\sqrt{n!}}\ket{n}
\label{eq:icbpsi}
\end{equation}
which gives rise to the form factors $\mathcal{F}_{\mathcal{B}}$ in Eq~\ref{eq:Funiform}.
This wavefunction can be thought of as a natural $S\rightarrow\infty$ limit of the finite-$S$ state, Eq~\ref{eq:vsigma}, with $n=S+\sigma$.
% Furthermore, the dispersion is independent of the spinors,  

The origin of Berry curvature in this model can be understood by viewing the Hamiltonian as a map from the 2D plane of $\bm{k}$ to the non-commutative phase space of $\bm{R}=(X,Y)$  defined by $X=(b+b^\dagger)/\sqrt{2}$, $Y=i(b-b^\dagger)/\sqrt{2}$ (in the same way that a two-band tight binding Hamiltonian is a map from the BZ torus to the Bloch sphere).
The state $\ket{s_{\bm{k}}^{\mathcal{B}}}$ is the coherent state centered at $\bm{R}=\sqrt{\mathcal{B}}\bm{k}$.
The Berry phase associated with a closed path in $\bm{k}$-space is equal to the total enclosed area swept out in $\bm{R}$-space.
The uniform Berry curvature $\mathcal{B}$ is then a direct consequence of the linear map $\bm{k}\rightarrow\bm{R}=\sqrt{\mathcal{B}}\bm{k}$.
% The uniform Berry curvature is the consequence of the non-commutativity $[X,Y]=-i$ and the linearity of the map $\bm{k}\rightarrow\bm{R}=\sqrt{\mathcal{B}}\bm{k}$.
The advantage of this model is that it works directly in the uniform Berry curvature limit, providing a concrete microscopic realization of the spinors $\ket{s_{\bm{k}}^{\mathcal{B}}}$.
}

\section{Periodic potential}\label{sec:periodic}
\begin{figure}[t]
\begin{center}
\includegraphics[width=0.48\textwidth]{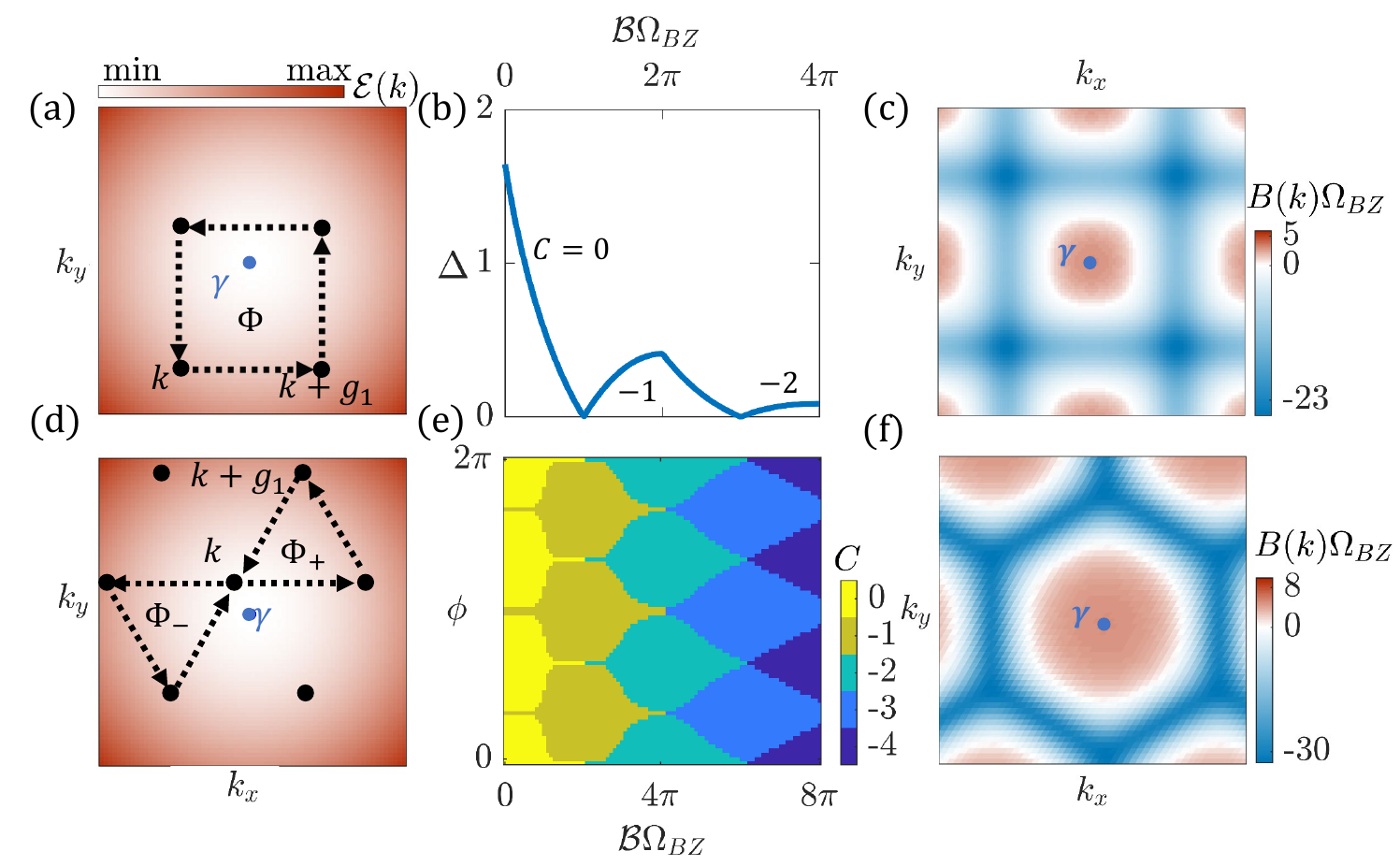}
\end{center}
\caption{ 
The ideal parent band in a periodic scalar potential.
For the $C_4$ symmetric potential,
(a) illustrates the effective momentum space square lattice Hofstadter model at each $\bm{k}$, (b) shows the gap between the first and second minibands, and the Chern number as a function of $\mathcal{B}$, and (c) shows the  Berry curvature distribution { of the first miniband (which differs from the parent Berry curvature near the zone edges where gaps are opened)}.
For the $C_3$ symmetric potential,
(d) shows the effective triangular lattice Hofstadter model, (e) shows the Chern number as a function of $\mathcal{B}$ and potential shape parameter $\phi$, and (f) shows the Berry curvature.
  Parameters used for (b,e) are $m=\frac{1}{2}$, $g=1$, and $U_0=1 $ ($U_0=5$) and
  (c,f) $\mathcal{B}\Omega_{\mathrm{BZ}}=2\pi$, $U_0=1$ ($U_0=0.5$,$\phi=\pi$), 
  for the $C_4$ ($C_3$) symmetric potential. 
 }
\label{fig:ideal}
\end{figure}

We now proceed with our analysis. 
It is useful to first consider the effect of the periodic potential $\hat{\mathcal{H}}_{\mathrm{pot}}$.
Let us focus first on the $C_4$ symmetric potential with period $a$, 
\begin{equation}
\begin{split}
U(\bm{r})=&-2U_0\left(\cos(\frac{2\pi x}{a})+\cos(\frac{2\pi y}{a})\right)\\
=& -U_0\sum_{j=1}^{4} e^{i \bm{b}_j\cdot \bm{r}}
\end{split}
\end{equation}
where $\bm{b}_j=g(\cos(\frac{\pi (j-1)}{2}),\sin(\frac{\pi (j-1)}{2}))$ are reciprocal lattice vectors, and $g=\frac{2\pi}{a}$.

The Hamiltonian takes the form $\hat{\mathcal{H}}=\hat{\mathcal{H}}_0+\hat{\mathcal{H}}_{\mathrm{pot}}$, with
\begin{equation}
\begin{split}
\hat{\mathcal{H}}_0=&\sum_{\bm{k},\bm{g}}\mathcal{E}(\bm{k}+\bm{g})c^\dagger_{\bm{k},\bm{g}}c_{\bm{k},\bm{g}}\\
\hat{\mathcal{H}}_{\mathrm{pot}}=&-U_0e^{-\frac{\mathcal{B}}{4}g^2}\sum_{\bm{k},\bm{g},j} c^\dagger_{\bm{k},\bm{g}+\bm{b}_j} e^{i\frac{\mathcal{B}}{2}(\bm{k}+\bm{g})\cross\bm{b}_j}c_{\bm{k},\bm{g}}
\end{split}
\end{equation}
where $c_{\bm{k},\bm{g}}=c_{\bm{k}+\bm{g}}$ and, from this point onwards, $\bm{k}$ are momenta within the first Brillouin zone (which remains a good quantum number), and $\bm{g}$ are reciprocal lattice vectors.
The periodic potential causes the original dispersion $\mathcal{E}(\bm{k})$ to split into multiple minibands.
The prospect of engineering topological minibands from parent bands with non-trivial topology has been recently explored~\cite{zeng2024gatetunable,tan2024designing,su2022massive,ghorashi2023topological,ghorashi2023multilayer,suri2023superlattice}.
In general, the miniband structure and topology depend crucially on the quantum geometry of the original band~\cite{liangprivate}.
% We henceforth refer to these minibands simply as ``bands''.

For each $\bm{k}$, $\hat{\mathcal{H}}=\hat{\mathcal{H}}_0+\hat{\mathcal{H}}_{\mathrm{pot}}$ can be mapped exactly to a momentum space version of the Hofstadter tight binding model ~\cite{Hofstadter1976Eneregy} on the square lattice of sites $\bm{g}_{n_1 n_2}=n_1\bm{b}_1+n_2\bm{b}_2$, ($n_1,n_2\in\mathbb{Z}$), with hopping amplitude $-U_0e^{-\frac{\mathcal{B}}{4}g^2}$, in the presence of a uniform a magnetic field $\mathcal{B}$ and a parabolic confining potential $\mathcal{E}(\bm{k}+\bm{g}_{n_1 n_2})$, as illustrated in Fig~\ref{fig:ideal}a.
This Hofstadter model has an effective magnetic flux of $\Phi=\mathcal{B}\Omega_{\mathrm{BZ}}$ through each plaquette, where $\Omega_{\mathrm{BZ}}=g^2$.
Crucially, the properties of the Hofstadter model depends only on the magnetic flux modulo $2\pi$.
This implies a relation between the Hamiltonian with $\mathcal{B}$ to that with $\mathcal{B}+2\pi/\Omega_{\mathrm{BZ}}$.
We now establish this periodicity, and use it to make exact statements about the topology of the resulting minibands.

Consider the unitary transformation 
\begin{equation}
\hat{W} c_{\bm{k},\bm{g}}^{\dagger[\mathcal{B}]} \hat{W}^{\dagger} = W_{\bm{g}}(\bm{k})c_{\bm{k},\bm{g}}^{\dagger[\mathcal{B}+2\pi/\Omega_{\mathrm{BZ}}]}
\end{equation}
which shifts $\mathcal{B}$ in the creation/annihilation operators, but more importantly applies a phase factor
\begin{equation}
W_{\bm{g}}(\bm{k})=\exp{i\pi\left(\frac{\bm{k}\cross\bm{g}}{\Omega_{\mathrm{BZ}}}+\omega(\bm{g})\right)}
\label{eq:Udef}
\end{equation}
where $\omega(\bm{g})=g_x g_y/g^2=n_1 n_2$.
This transformation leaves the kinetic term $\hat{\mathcal{H}}_0$ invariant (it affects the microscopic wavefunctions created by $c^\dagger_{\bm{k}}$, but the form of the second quantized Hamiltonian is unchanged).

The reason for the phase factors is so that $\hat{W}$ will transform $\hat{\mathcal{H}}_{\mathrm{pot}}$,
 when viewed as a function of $\mathcal{B}$ and $U_0$, according to
\begin{equation}
  \hat{\mathcal{H}}_{\mathrm{pot}}[\mathcal{B}+2\pi/\Omega_{\mathrm{BZ}},U_0] = \hat{W}\hat{\mathcal{H}}_{\mathrm{pot}}[\mathcal{B},U_0e^{-\frac{\pi}{2}}] \hat{W}^{\dagger} 
  \label{eq:HamPeriodicity}
 \end{equation}
Thus implying that the total spectrum of $\hat{\mathcal{H}}$ is periodic under a shift of $\mathcal{B}$ by $2\pi/\Omega_{\mathrm{BZ}}$, provided we also rescale the potential strength $U_0$.
The phase factors $W_{\bm{g}}(\bm{k})$ can be interpreted as the ``gauge transformation'' which adds $2\pi$ flux to the momentum space Hofstadter model, and the rescaling of $U_0$ corrects for the quantum distance factor in $\mathcal{F}$.

Despite the spectrum being periodic, we now demonstrate that topology is not.  
Specifically, 
\begin{equation}
C[\mathcal{B}+2\pi/\Omega_{\mathrm{BZ}},U_0] = C[\mathcal{B},U_0e^{-\frac{\pi}{2}}]-1 
\label{eq:ChernPeriodicity}
\end{equation}
where $C$ is the Chern number of (any) isolated miniband.
Thus, counter-intuitively, an \emph{increase} of the parent band Berry curvature by $2\pi/\Omega_{\mathrm{BZ}}$ results in a \emph{decrease} of the miniband Chern number by $1$.

To prove Eq~\ref{eq:ChernPeriodicity}, we consider the first quantized wavefunction of $\hat{\mathcal{H}}[\mathcal{B}_0,U_0e^{-\frac{\pi}{2}}]$ in a single isolated band, $|\psi_{\bm{k}}^{(0)}\rangle=\sum_{\bm{g}}\psi_{\bm{k},\bm{g}}^{(0)}e^{i(\bm{k}+\bm{g})\cdot \bm{r}}|s^{\mathcal{B}_0}_{\bm{k}+\bm{g}}\rangle $.
As a result of Eq~\ref{eq:HamPeriodicity}, the wavefunction
\begin{equation}
|\psi^{(1)}_{\bm{k}}\rangle = \sum_{\bm{g}}
W_{\bm{g}}(\bm{k})\psi^{(0)}_{\bm{k},\bm{g}}  e^{i(\bm{k}+\bm{g})\cdot \bm{r}}|s_{\bm{k}+\bm{g}}^{\mathcal{B}_1}\rangle
\end{equation}
is the corresponding eigenstate of $\mathcal{H}[\mathcal{B}_1,U_0]$, where $\mathcal{B}_1=\mathcal{B}_0+2\pi/\Omega_{\mathrm{BZ}}$.
We use the superscripts $(0)$ to denote things relating to the $\mathcal{B}_0$ model, and $(1)$ to denote those relating to $\mathcal{B}_1$.
The Berry curvature 
is calculated from the cell-periodic part of the wavefunction, $|u_{\bm{k}}^{(1)}\rangle=e^{-i\bm{k}\cdot\bm{r}}|\psi_{\bm{k}}^{(1)}\rangle$,
which gives rise to
\begin{equation}
B^{(1)}(\bm{k}) = B^{\prime}(\bm{k}) + \frac{\pi}{\Omega_{\mathrm{BZ}}}\sum_{\bm{g},\mu}g_{\mu}\partial_{k_\mu}|\psi^{(0)}_{\bm{g}}(\bm{k})|^2
\label{eq:B1}
\end{equation}
where $B^{\prime}$ is the Berry curvature computed from $|u_{\bm{k}}^\prime\rangle=\sum_{\bm{g}}\psi_{\bm{k},\bm{g}}^{(0)}e^{i\bm{g}\cdot\bm{r}}|s_{\bm{k}+\bm{g}}^{\mathcal{B}_1}\rangle$.
It can be shown that $\int_{\mathrm{BZ}}B^\prime(\bm{k})d\bm{k}=\int_{\mathrm{BZ}}B^{(0)}(\bm{k})d\bm{k}=2\pi C^{(0)}$ gives the original Chern number.
The final Chern number is therefore
\begin{equation}
C^{(1)}=\frac{1}{2\pi}\int_{\mathrm{BZ}} B^{(1)}(\bm{k})d\bm{k} = C^{(0)}-1
\end{equation}
where the $-1$ is due to second term in Eq~\ref{eq:B1}, which
originates from the $W_{\bm{g}}(\bm{k})$ factor in $|\psi^{(1)}\rangle$.
This result follows by integrating, using $|\psi^{(0)}_{\bm{k}+\bm{g}^\prime,\bm{g}}|=|\psi^{(0)}_{\bm{k},\bm{g}+\bm{g}^\prime}|$, and performing the $\bm{g}$ summation by parts, 
thus proving Eq~\ref{eq:ChernPeriodicity}.

Fig~\ref{fig:ideal}b shows the gap and Chern number of the first band, calculated numerically, as a function of $\mathcal{B}$, confirming this result.
Fig~\ref{fig:ideal}c shows the Berry curvature of the first miniband, demonstrating this counter-intuitive feature.  
Near the $\gamma$ point of the BZ, $B(\bm{k})$ is positive (inherited from the positive parent Berry curvature), but due to superlattice effects, a gap is opened up near the zone boundaries that give rise to negative $B(\bm{k})$, precisely overcancelling the parent Berry curvature to give a total negative Chern number.

We can also consider a $C_3$ symmetric potential, 
\begin{equation}
U(\bm{r})=-\sum_{j=1,3,5} U_0 e^{i\phi}e^{i\bm{b}_j\cdot \bm{r}} + c.c.
\end{equation}
where $\bm{b}_j=g(\cos(\frac{\pi (j-1)}{3}),\sin(\frac{\pi (j-1)}{3}))$ and $\phi$ is a parameter that controls the shape of the potential.
There are a few minor differences from the $C_4$ symmetric potential.
First, the Hamiltonian maps to a momentum space Hofstadter model on the triangular lattice $\bm{g}_{n_1 n_2}=n_1 \bm{b}_1+n_2\bm{b}_2$, instead of the square lattice, with the same confining parabolic potential (Fig~\ref{fig:ideal}d).
The hopping amplitudes are $-U_0e^{-\frac{\mathcal{B}}{4}g^2\pm i\phi}$ which, in addition to the uniform $\mathcal{B}$, give rise to an effective (staggered) magnetic flux of $\Phi_{\pm}=\frac{1}{2}\mathcal{B}\Omega_{\mathrm{BZ}}\pm3\phi$ on each plaquette, where $\Omega_{\mathrm{BZ}}=\sqrt{3}g^2/2$ (each triangular plaquette has an area of $\Omega_{\mathrm{BZ}}/2$).
The triangular lattice Hofstadter model comes back to itself with the addition of $\pi$ flux per plaquette, but with an opposite sign of hopping amplitudes.  
This manifests as the periodicity relation
\begin{equation}
  \hat{\mathcal{H}}_{\mathrm{pot}}[\mathcal{B}_0+2\pi/\Omega_{\mathrm{BZ}},U_0 ] = \hat{W}\hat{\mathcal{H}}_{\mathrm{pot}}[\mathcal{B}_0,-U_0e^{-\frac{\pi}{\sqrt{3}}}] \hat{W}^{\dagger} 
  \label{eq:HamPeriodicityTri}
 \end{equation}
 with $W$ given as in Eq~\ref{eq:Udef}, except with $\omega(\bm{g})=(n_1-1)(n_2-1)$.
 Provided that the spectrum is gapped so Chern number is well defined, 
\begin{equation}
C[\mathcal{B}_0+2\pi/\Omega_{\mathrm{BZ}},U_0] = C[\mathcal{B}_0,-U_0 e^{-\frac{\pi}{\sqrt{3}}}]-1 
\label{eq:ChernPeriodicityTri}
\end{equation}
obeys a similar periodicity relation.
Fig~\ref{fig:ideal}(e,f) shows the phase diagrams and Berry curvature for the $C_3$ symmetric potential.

A consequence of these results is that, for $\mathcal{B}=2\pi N/\Omega_{\mathrm{BZ}}$, $N\in\mathbb{Z}$, each isolated miniband must carry Chern number $C=-N$.
This follows from both Eq~\ref{eq:ChernPeriodicity} or Eq~\ref{eq:ChernPeriodicityTri} and the fact that the Chern number must be zero for $\mathcal{B}=0$ where there is no source of time-reversal symmetry breaking.
This Chern number is equal in magnitude, but opposite in sign, to the integrated parent Berry curvature over the BZ.
This is true even in the limit of an infinitely strong potential.  

\section{Strong interactions}\label{sec:interaction}

We now consider the effect of interactions $\hat{\mathcal{H}}_{\mathrm{int}}$, in the absence of a periodic potential so continuous translation symmetry is present.
In a topologically trivial parent band ($\mathcal{B}=0$), strong interactions can lead to spontaneous breaking of continuous translation symmetry in the form of a Wigner crystal.
As we shall demonstrate, strong interactions will also lead to spontaneously broken continuous translation symmetry in the presence of strong parent Berry curvature, but with non-zero Chern number, thus realizing an AHC.

In general, the periodicity in $\mathcal{B}$ is no longer an exact feature of the interacting problem.  
This is because the interactions involve quartic terms from different momenta $\bm{k}$, effectively coupling the momentum space Hofstadter models from different $\bm{k}$.
Nevertheless, we show that there still remains a remnant of this periodicity, and that it results in topological states with the \emph{same} Chern number as the parent Berry curvature $\mathcal{B}$ (opposite to our conclusions for periodic potential).

We consider the Hamiltonian $\hat{\mathcal{H}}_0+\hat{\mathcal{H}}_{\mathrm{int}}$, where $\hat{\mathcal{H}}_{\mathrm{int}}$ is given in Eq~\ref{eq:Hint}.
We allow for continuous translation symmetry breaking down to a discrete unit cell with reciprocal lattice vectors $\bm{g}$ with $C_3$ symmetry.
We assume the period is determined by the density of electrons such that there is one electron per unit cell, $\Omega_{\mathrm{BZ}}=4\pi^2 n$ where $n$ is the electron density, as expected for a Wigner crystal.
We consider decomposing the interaction into two terms, the Hartree term in which $\bm{k}_1=\bm{k}_4$, $\bm{k}_2=\bm{k}_3$, and the Fock term in which $\bm{k}_1=\bm{k}_3$, $\bm{k}_2=\bm{k}_4$ (within the BZ).
That is,
$\hat{\mathcal{H}}_{\mathrm{int}}\approx \hat{\mathcal{H}}_{\mathrm{H}}+\hat{\mathcal{H}}_{\mathrm{F}}$,
\begin{equation}
\begin{split}
\hat{\mathcal{H}}_{\mathrm{H}}=\frac{1}{2A}\sum_{\substack{\bm{k}_1\bm{k}_2\\\bm{g}_1\bm{g}_2\bm{g}_3\bm{g}_4}} {V}^{\bm{k}_1\bm{k}_2\bm{k}_2\bm{k}_1}_{\bm{g}_1\bm{g}_2\bm{g}_3\bm{g}_4}
({c}_{\bm{k}_1\bm{g}_1}^\dagger c_{\bm{k}_1\bm{g}_4}) ({c}_{\bm{k}_2\bm{g}_2}^\dagger  {c}_{\bm{k}_2\bm{g}_3})\\
\hat{\mathcal{H}}_{\mathrm{F}}=-\frac{1}{2A}\sum_{\substack{\bm{k}_1 \bm{k}_2\\\bm{g}_1\bm{g}_2\bm{g}_3\bm{g}_4}} {V}^{\bm{k}_1\bm{k}_2\bm{k}_1\bm{k}_2}_{\bm{g}_1\bm{g}_2\bm{g}_3\bm{g}_4}
({c}_{\bm{k}_1\bm{g}_1}^\dagger c_{\bm{k}_1\bm{g}_3}) ({c}_{\bm{k}_2\bm{g}_2}^\dagger  {c}_{\bm{k}_2\bm{g}_4})
\end{split}
\label{eq:HHF}
\end{equation}
where
\begin{equation}
\begin{split}
{V}^{\bm{k}_1\bm{k}_2\bm{k}_3\bm{k}_4}_{\bm{g}_1\bm{g}_2\bm{g}_3\bm{g}_4} =& V(\bm{k}_1+\bm{g}_1-\bm{k}_4-\bm{g}_4)
\mathcal{F}_{\mathcal{B}}(\bm{k}_1+\bm{g}_1,\bm{k}_4+\bm{g}_4)\\
&\times \mathcal{F}_{\mathcal{B}}(\bm{k}_2+\bm{g}_2,\bm{k}_3+\bm{g}_3) \times \delta(\dots)
\end{split}
\end{equation}
and the $\delta(\dots)$ ensures total momentum conservation.
Both $\hat{\mathcal{H}}_{\mathrm{H}}$ and $\hat{\mathcal{H}}_{\mathrm{F}}$ are, at this point, still fully interacting.
The HF approximation follows from a mean-field treatment of the interactions, $(c^\dagger c)(c^\dagger c)\approx \langle c^\dagger c\rangle c^\dagger c + c^\dagger c \langle c^\dagger c\rangle - \langle c^\dagger c\rangle\langle c^\dagger c\rangle$, in which $\hat{\mathcal{H}}_{\mathrm{H,F}}$ give rise to the Hartree  and Fock terms respectively, and then solving the set of equations self-consistently.

Let us again consider the unitary transformation $\hat{W}$ from Eq~\ref{eq:HamPeriodicityTri}.
Indeed, somewhat remarkably, $\hat{\mathcal{H}}_\mathrm{H}$, viewed as a functional of $\mathcal{B}$ and the Fourier transform of the density-density interaction $V(q)$, satisfies the exact periodicity
\begin{equation}
\hat{\mathcal{H}}_{\mathrm{H}}[\mathcal{B}+2\pi/\Omega_{BZ},V(q)]
=\hat{W}\hat{\mathcal{H}}_{\mathrm{H}}[\mathcal{B},V(q)e^{-\frac{\pi q^2}{\Omega_{\mathrm{BZ}}}}] \hat{W}^{\dagger}
\end{equation}
However, the same does not apply to the $\mathcal{H}_{\mathrm{F}}$.
Instead, 
\begin{equation}
\hat{\mathcal{H}}_{\mathrm{F}}[\mathcal{B}+2\pi/\Omega_{BZ},V(q)]
=\hat{Q}\hat{\mathcal{H}}_{\mathrm{F}}[\mathcal{B},V(q)e^{-\frac{\pi q^2}{\Omega_{\mathrm{BZ}}}}] \hat{Q}^\dagger
\label{eq:FockPeriodicity}
\end{equation}
where $\hat{Q}$ is defined 
\begin{equation}
\hat{Q} c_{\bm{k},\bm{g}}^{\dagger[\mathcal{B}]} \hat{Q}^{\dagger} = W_{\bm{g}}^{*}(\bm{k})c_{\bm{k},\bm{g}}^{\dagger[\mathcal{B}+2\pi/\Omega_{\mathrm{BZ}}]}
\end{equation}
Thus, there is not a single unitary transformation that accomplishes the mapping for both terms simultaneously.

The presence of this exact mapping of the quartic Hamiltonians $\hat{\mathcal{H}}_{\mathrm{H}}$ and $\hat{\mathcal{H}}_{\mathrm{F}}$ individually is quite remarkable.
It is especially interesting considering that, in the periodic potential case, the momentum space Hofstadter model fails to exhibit the exact $2\pi$ flux-per-plaquette periodicity when generic longer-range hoppings are included (that is, the $\hat{\mathcal{H}}_{\mathrm{pot}}$ we consider only satisfies the exact $2\pi/\Omega_{\mathrm{BZ}}$ periodicity because the potential $U(\bm{r})$ is taken to consist of first harmonic terms $\bm{b}_j$).
Nevertheless, both $\hat{\mathcal{H}}_{\mathrm{H}}$ and $\hat{\mathcal{H}}_{\mathrm{F}}$, which in principle contains correlated hopping terms of all ranges, enjoys this exact mapping without restriction.

We now argue that the Fock term will dominate when $\mathcal{B}$ is significant.  
Suppose $\mathcal{B}=2\pi N/\Omega_{\mathrm{BZ}}$.  
By repeated applications of $\hat{Q}^\dagger$, $\hat{\mathcal{H}}_{\mathrm{H}}+\hat{\mathcal{H}}_{\mathrm{F}}$ can be related to a zero-$\mathcal{B}$ Hamiltonian 
\begin{equation}
\begin{split}
\hat{Q}^{\dagger N}&(\hat{\mathcal{H}}_{\mathrm{H}}[\mathcal{B},V(q)]+\hat{\mathcal{H}}_{\mathrm{F}}[\mathcal{B},V(q)])\hat{Q}^{ N} = \\
&(\hat{Q}^{\dagger}\hat{W})^N
\hat{\mathcal{H}}_{\mathrm{H}}[0,\tilde{V}(q)] (\hat{W}^{\dagger}\hat{Q})^{ N}
+\hat{\mathcal{H}}_{\mathrm{F}}[0,\tilde{V}(q)]
\end{split}
\end{equation}
where $\tilde{V}(q)=V(q)e^{-\frac{\pi N}{\Omega_{\mathrm{BZ}}}q^2}$
is an effective interaction that is suppressed at large $q$ by a Gaussian factor.
For bare Coulomb interactions, the effective interaction takes the form $\tilde{V}(r)\sim e^{-\frac{\Omega_{\mathrm{BZ}}r^2}{8\pi N}}I_0(\frac{\Omega_{\mathrm{BZ}}r^2}{8\pi N})$ that is smoothed off at short distances. 
This Hamiltonian differs from a true trivial parent band model only in the presence of various phase factors in $\hat{\mathcal{H}}_{\mathrm{H}}$ introduced by the $\hat{W}^\dagger\hat{Q}$.
However, since $\hat{\mathcal{H}}_{\mathrm{H}}$ only involves momentum transfer $q$ with a minimum of $q=g$
(the $\tilde{V}(q=0)$ term is just the total charging energy), its magnitude is more strongly suppressed by the Gaussian factor than 
 $\hat{\mathcal{H}}_{\mathrm{F}}$, which also involves $\tilde{V}(q)$ with $0<q<g$.
As a result, $\hat{\mathcal{H}}_{\mathrm{H}}$ can be neglected for large enough  $\mathcal{B}$.
% {\color{red} We have also numerically verified that this is the case even for $\mathcal{B}=2\pi/\Omega_{\mathrm{BZ}}$ for Coulomb interactions~\cite{supplement}.}
Neglecting $\hat{\mathcal{H}}_{\mathrm{H}}$, the mapping in Eq~\ref{eq:FockPeriodicity} applies to the fully interacting Hamiltonian $\hat{\mathcal{H}}_0+\mathcal{H}_{\mathrm{F}}$ exactly.

Now, suppose we have a self-consistent solution to the HF mean field Hamiltonian at $\mathcal{B}=2\pi N/\Omega_{\mathrm{BZ}}$, with sufficiently strong interactions to create a charge gap, where the first quasiparticle band is filled.
In the Fock dominated regime, the Hamiltonian is unitarily related to a zero-$\mathcal{B}$ problem with an effective interaction $\tilde{V}(q)$.  
The corresponding zero-$\mathcal{B}$ state describes a Wigner crystal with a filled quasiparticle band of trivial topology.
By the same argument for the change in topology in Eq~\ref{eq:ChernPeriodicity}, except with the phase factors ${W}_{\bm{g}}^*(\bm{k})$, we conclude that the filled band of the $\mathcal{B}=2\pi N/\Omega_{\mathrm{BZ}}$ model must have a Chern number of $C=+N$, i.e. it describes an AHC.  

Note that the Chern number is positive, which contrasts to the negative Chern band in the presence of a strong periodic potential.
This indicates that the two limits, strong potential versus strong interactions, are distinct and, when both present, will generically compete with one another.
This result is unexpected, as the two effects are typically complementary in a topologically trivial band.

\begin{figure}[t]
\begin{center}
\includegraphics[width=0.48\textwidth]{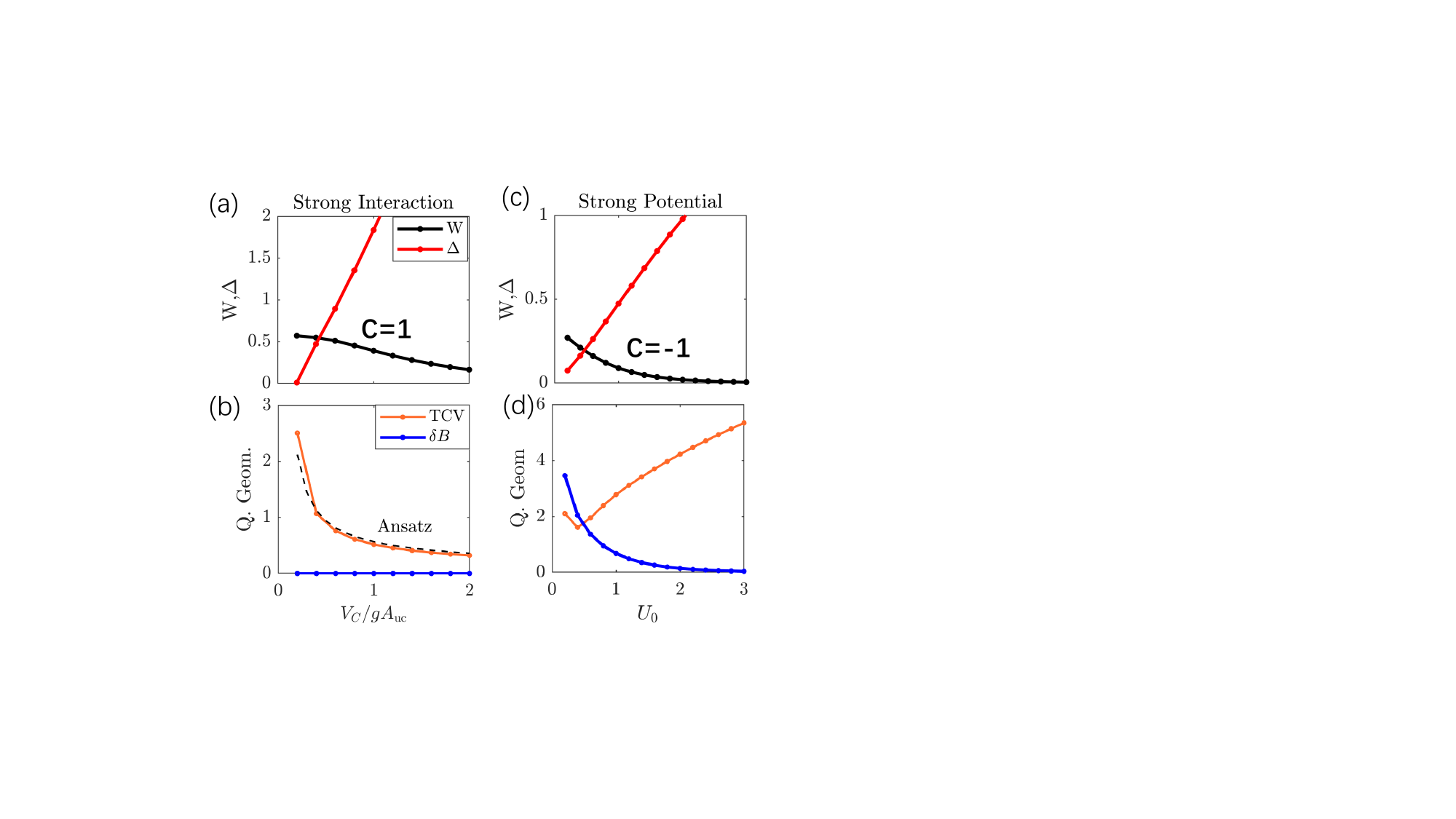}
\end{center}
\caption{
Energetic and quantum geometric properties of the model. (a) The charge gap $\Delta$ and bandwidth $W$ of the filled band obtained from solving the mean-field HF Hamiltonian self-consistently, as a function of Coulomb interaction strength $V_C$.
(b) The Berry curvature fluctuation ($\delta B$) and trace condition violation (TCV) of the filled band are shown.
The dashed line shows the TCV from the analytic calculation based on the wavefunction ansatz.
(c,d) The corresponding quantities for a non-interacting model with a honeycomb lattice periodic potential $U_0$, $\phi=\pi$.
{ Note that TCV does not vanish even as the gap approaches zero due to contributions at the Brillouin zone boundary. }
All energies are measured in units of ${g^2}/{(2m)}$.
% {\color{blue}change "Analytic" to "Ansatz", capitalize the $c$ in $V_c\rightarrow V_C$, fix font $A_{uc}\rightarrow A_{\mathrm{uc}}$, and change the fraction style $\frac{V_C}{g A_{\mathrm{uc}}}\rightarrow V_C/(g A_{\mathrm{uc}})$.}
}
\label{fig:strongint}
\end{figure}
To verify our analysis, we numerically solve the full self-consistent HF equations in Fig~\ref{fig:strongint}.
We take the Coulomb interaction $V(q)=V_C/q$, with no periodic potential, and focus on $\mathcal{B}=2\pi/\Omega_{\mathrm{BZ}}$.
The HF mean-field Hamiltonian is solved self-consistently at filling $\nu=1$ electron per unit cell.
The calculation is performed by keeping states up to a cutoff $|\bm{g}|<\Lambda$.
Fig~\ref{fig:strongint}a shows the bandwidth of the filled quasiparticle band $W$, and charge gap $\Delta$, as a function of interaction strength $V_C$. 
Above a critical $V_C$, $\Delta$ becomes non-zero signaling a transition into a crystalline state with spontaneously broken continuous translation symmetry, while the bandwidth $W$ becomes small.
Numerical integration of the Berry curvature, properly taking into account the wavefunction overlaps of the parent band, reveals a Chern number $C=1$, in agreement with the above analysis in the Fock-dominated regime.

To contrast, we show in
Fig~\ref{fig:strongint}c the analogous results for a non-interacting Hamiltonian ($V_C=0$), in the $C_3$ symmetric potential $U(\bm{r})$ at the same $\mathcal{B}=2\pi/\Omega_{\mathrm{BZ}}$.
{The potential is chosen to be honeycomb ($\phi=\pi$) as it maximizes the gap (it maps to a zero-$\mathcal{B}$ problem in a triangular lattice potential).}
For any $U_0>0$, the gap $\Delta>0$, and bandwidth $W$ becomes very small with increasing $U_0$.
Integration of the Berry curvature reveals that the first band is topological with $C=-1$, as expected. 

We now turn to the quantum geometric properties of the resulting bands.
We consider two quantum geometric indicators that have been proposed to probe the suitability of a band to realizing fractionalized phases at partial filling.
We consider the Berry curvature fluctuation
\begin{equation}
(\delta B)^2 = \frac{\Omega_{\mathrm{BZ}}}{4\pi^2}\int_{\mathrm{BZ}} (B(\bm{k})-\bar{B})^2 d\bm{k}
\end{equation}
where $\bar{B}=\Omega_{\mathrm{BZ}}^{-1}\int_{\mathrm{BZ}}B(\bm{k})d\bm{k}$,
and the average trace condition violation
\begin{equation}
\mathrm{TCV}=\frac{1}{2\pi}\int_{\mathrm{BZ}} \left(\Tr[g^{FS}_{\mu\nu}(\bm{k})]-|B(\bm{k})|\right)d\bm{k}
\end{equation}
which measures the amount by which the trace inequality, $\Tr[g^{\mathrm{FS}}_{\mu\nu}(\bm{k})]\geq|B(\bm{k})|$, is exceeded.
A band with $\mathrm{TCV}=0$ is said to be ``ideal'' for the realization of fractional Chern insulators ~\cite{wang2021exact,ledwith2022vortexability,claassen2015position,Ledwith2020fractionalchern,Lee2017engineering,jackson2015geometric,BrunoKahler2021}.

We show $\delta B$ and TCV for the filled HF band of the interacting model in Fig~\ref{fig:strongint}b, and the corresponding plot for the first band in a the periodic potential without interaction in Fig~\ref{fig:strongint}d.
The Berry curvature distribution for both cases becomes increasingly uniform with $V_C$ or $U_0$.
However, they differ strikingly in terms of their TCV.
While the interacting $C=1$ AHC becomes increasingly ideal with strong interactions, the  non-interacting $C=-1$ CI with a periodic potential is the opposite.
This striking difference will be explained in the next section, based on an analytic ansatz for the wavefunctions of these two states.

\section{Wavefunction ansatz}

\begin{figure}[t]
\begin{center}
\includegraphics[width=0.48\textwidth]{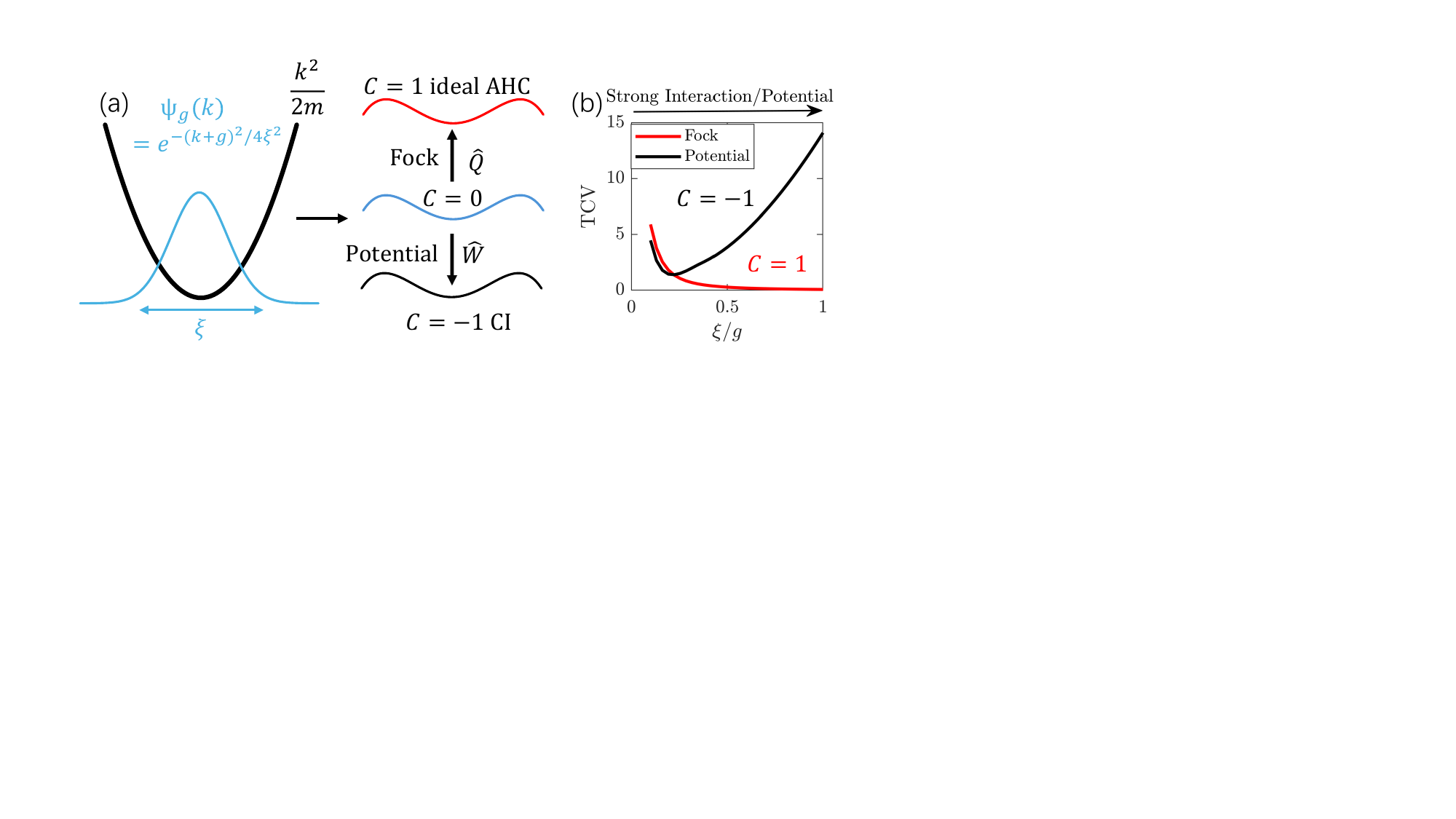}
\end{center}
\caption{
Wavefunction ansatz and ideal AHC.  (a) Illustration of the wavefunction for the trivial $C=0$ state in the $\mathcal{B}=0$ parent band, where $\xi$ is a tuning parameter; $\xi\rightarrow\infty$ describes the limit of strong interaction or potential.
By utilizing the transformation $\hat{Q}$ ($\hat{W}$), we arrive at a $C=1$ AHC ($C=-1$ CI) state for the $\mathcal{B}=2\pi/\Omega_{\mathrm{BZ}}$ model in the Fock (potential) dominated regime, respectively.
(b) The trace condition violation (TCV) is shown as a function of $\xi$, demonstrating that the AHC becomes ideal as $\xi\rightarrow\infty$. 
% {\color{blue}Remove the word ``non-ideal'', just say ``$C=-1$ CI''.}
}
\label{fig:ansatz}
\end{figure}

The mappings established in Sec~\ref{sec:periodic} and \ref{sec:interaction} are incredibly powerful.
They manage to relate the ground states of topological quantum systems, the $C=N$ AHC or the $C=-N$ Chern insulator, 
to those of a topologically trivial Wigner crystal or band insulator, which can essentially be described classically.
In this section, we leverage this relation to write down analytic wavefunctions based on a simple Gaussian ansatz for the trivial band insulator.
Remarkably, this analytic expression fully captures and explains the emergence of the ideal HF band of the strongly interacting AHC.

First, let us consider the ground state wavefunction of the zero-$\mathcal{B}$ Hamiltonian in an infinitely strong periodic potential (with a single potential minimum per unit cell).
{
This limit is equivalent to an array of harmonic oscillators localized at each potential minimum $\bm{R}$, each of which are approximately described by the Gaussian wavefunction $\phi_{\bm{R}}(\bm{r})\sim \exp{-\alpha |\bm{r}-\bm{R}|^2}$.
Orthogonal Bloch wavefunctions at crystal momentum $\bm{k}$ can be constructed from these states as 
$\psi_{\bm{k}}(\bm{r})\sim\sum_{\bm{R}}e^{i\bm{k}\cdot\bm{R}}\phi_{\bm{R}}(\bm{r})$.
This state has the Fourier representation 
$\psi_{\bm{k}}(\bm{r})=\sum_{\bm{g}}\psi_{\bm{k},\bm{g}}e^{i(\bm{k}+\bm{g})\cdot\bm{r}}$, with $\psi_{\bm{k},\bm{g}}=\mathcal{N}_{\bm{k}}e^{-\frac{(\bm{k}+\bm{g})^2}{4\xi^2}}$, 
where $\xi=\sqrt{\alpha}$ is a momentum space localization length and $\mathcal{N}_{\bm{k}}$ is a normalization constant, illustrated in Fig~\ref{fig:ansatz}a.
}
% The equivalent momentum space picture is a single isolated flat band, with the wavefunctions described by
% $ \psi_{\bm{g}}^{(0)}(\bm{k}) = \mathcal{N}_{\bm{k}} e^{-\frac{(\bm{k}+\bm{g})^2}{4\xi^2}} $
% where $\xi$ is a momentum space localization length, and $\mathcal{N}_{\bm{k}}$ is a normalization constant, as illustrated in Fig~\ref{fig:ansatz}a.
Although we have considered a strong periodic potential, the ground state in the presence of an infinitely strong repulsive interaction is a classical Wigner crystal, which can also be well described by a Slater determinant of such states.
We take this wavefunction as an ansatz and treat $\xi$, the momentum space localization length, as an effective parameter reflecting the overall strength of the interaction or periodic potential.
This ansatz becomes exact in the classical limit in which $\xi\rightarrow\infty$.

Based on this, we can write down the corresponding wavefunctions for the $\mathcal{B}=2\pi/\Omega_{BZ}$ Hamiltonian.
These follow from applications of either $\hat{W}$ or $\hat{Q}$.
The wavefunctions are
\begin{equation}
|\psi_{\bm{k}}^{\pm}\rangle=\mathcal{N}_{\bm{k}}\sum_{\bm{g}}e^{-\frac{(\bm{k}+\bm{g})^2}{4\xi^2}\pm i\pi\left[\frac{\bm{k}\cross\bm{g}}{\Omega_{\mathrm{BZ}}}+\omega(\bm{g})\right]}e^{i(\bm{k}+\bm{g})\cdot\bm{r}}|s_{\bm{k}+\bm{g}}^{2\pi/\Omega_{\mathrm{BZ}}}\rangle
\label{eq:ansatzpsipm}
\end{equation}
where the $+$ ($-$) corresponds to the case of strong potential (interaction).
The full many-body state will be a Slater determinant of $|\psi_{\bm{k}}^{\pm}\rangle$.

We directly compute the TCV of the bands described by $|\psi_{\bm{k}}^{\pm}\rangle$
in Fig~\ref{fig:ansatz}b, as a function of $\xi$.
The behavior is similar to that observed numerically from HF in Fig~\ref{fig:strongint}(b,d).
Namely, in the strong potential limit $|\psi_{\bm{k}}^{+}\rangle$, TCV increases with increasing $\xi$, while in the strong interacting limit $|\psi_{\bm{k}}^{-}\rangle$, TCV decreases with increasing $\xi$ becoming more ideal.
For large $\xi$, the trace of the Fubini-Study metric (and the Berry curvature) is $\bm{k}$-independent, and an analytic expression can be computed directly from $|\psi^{\pm}_{\bm{k}}\rangle$,
\begin{equation}
\Tr[g_{\mu\nu}^{\mathrm{FS}(\pm)}]=\frac{2\pi}{\Omega_{\mathrm{BZ}}}+ (1\pm 1)\frac{4\pi^2\xi^2}{\Omega_{\mathrm{BZ}}^2} + \frac{1}{2\xi^2}+ \mathcal{O}(e^{-\frac{2\pi^2\xi^2}{\Omega_{\mathrm{BZ}}}})
\label{eq:ansatztrace}
\end{equation}
so that, in the Fock-dominated AHC, the $\xi^2$ term is zero and $\Tr [g]-|B|\rightarrow 0$ becomes perfectly ideal at large $\xi$,
in good agreement  with Fig~\ref{fig:ansatz}b.
Thus, interactions separate out an ideal AHC Chern band from the parent band (reminiscent of how ideal higher Chern bands can be decomposed into ideal Chern 1 bands~\cite{JKDong2023PRR}).
{ This suggests an interpretation of the AHC as a generalized quantum Hall ferromagnet.}
% {\color{red} This suggests an interpretation of the AHC in the $m\rightarrow\infty$ (or $V_C\rightarrow\infty$) limit as a generalized quantum Hall ferromagnet.
% {In this picture, the ideal ``$C=\infty$'' flat parent band, folded in the BZ, is reinterpreted as an infinite number of $C=1$ Landau levels. Spontaneously polarizing into one of these Landau levels, as in a quantum Hall ferromagnet, breaks translation symmetry and results in the AHC state.}

%{\color{red}
%One way to understand the emergence of an ideal Chern band is that the Fock energy favors a state with minimized quantum metric.
%}

{
% To determine the quantitative agreement of the ansatz wavefunction with HF, we now obtain $\xi$ as a function of model parameters.
We now treat this wavefunction as a variational ansatz, and minimize the total energy with respect to $\xi$.
We consider the case of the AHC with Coulomb interactions.
Empirically, we find that the contribution of the Hartree term to the total energy in our HF calculations is orders of magnitude smaller than the kinetic and Fock terms~\cite{supplement}, we therefore neglect it here.
The total (kinetic and Fock) energy per particle at large $\xi$ can then obtained by utilizing the unitary mapping to the trivial parent band~\cite{supplement},
\begin{equation}
E_{\mathrm{AHC}}(\xi) \approx \frac{\xi^2}{m}-\frac{V_C\xi/(4\sqrt{\pi})}{\sqrt{1+\frac{4\pi\xi^2}{\Omega_{\mathrm{BZ}}}}}
\end{equation}
which can be minimized to obtain the optimal $\xi$ as a function of $V_C$, $\xi_{\mathrm{opt}}(V_C)$.
The TCV as a function of $V_C$ can then be obtained from Eq~\ref{eq:ansatztrace} and assuming constant $|B|=2\pi/\Omega_{\mathrm{BZ}}$, resulting in $\mathrm{TCV}\approx 1/(2\xi^2_{\mathrm{opt}}(V_C))$.
The result of this (purely analytic) calculation is shown in Fig~\ref{fig:strongint}b, which demonstrates striking quantitative agreement with HF numerics.
% For large $V_C$, this gives a scaling of $\mathrm{TCV}\approx 4\pi(V_C m\Omega_{\mathrm{BZ}}^{\frac{3}{4}})^{-\frac{1}{2}}$.

% For a periodic potential $U_0$, $\xi$ is determined by minimizing the energy of the Gaussian wavepacket expanded about the effective potential minima
% % To test the quantitative agreement of this ansatz wavefunction, we must determine $\xi$ 

% Discussion of Hartree, Fock, and kinetic energy, and analytic expression for determining $\xi$.
}

\begin{figure}[t]
\begin{center}
\includegraphics[width=0.48\textwidth]{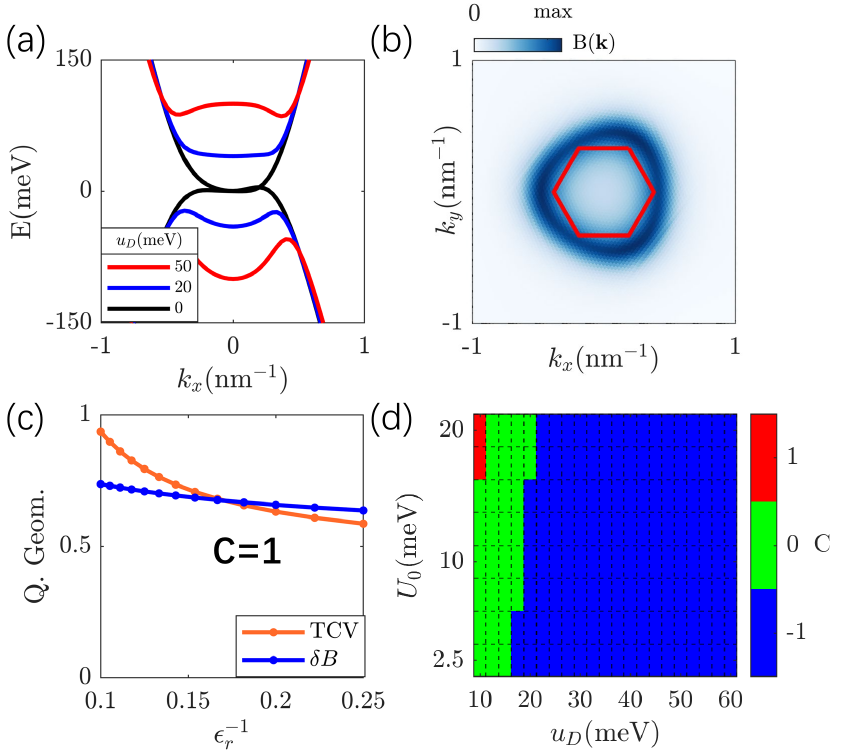}
\end{center}
\caption{
{Rhombohedral pentalayer graphene (R5G).
(a) The parent band structure near $K$ is shown for various choices of displacement field, modeled as a layer potential $u_D$.
(b) We focus on a single spin-valley component with positive parent Berry curvature, shown for displacement field $u_D=\SI{50}{meV}$, with the imposed Brillouin zone shown.
% (b) 
% The parent Berry curvature distribution is positive, shown for $u_D=\SI{50}{meV}$, and the imposed Brillouin zone is shown in red.
(c) With gate-screened Coulomb interactions at $\nu=1$, the HF ground state is a $C=1$ AHC. 
The quantum geometric properties of the HF band is shown as a function of the dielectric constant. (d) The topological phase diagram in the presence of a periodic potential as a function of potential strength $U_0$ and $u_D$, which instead shows a dominant $C=-1$ phase.
Model details are available in the supplemental material~\cite{supplement}.}
% {\color{blue}(fix caption once figure is updated)}
}
\label{fig:pentalayer}
\end{figure}

\section{Beyond the ideal parent band}
% \subsection{Beyond the ideal parent band}

{
We have presented several unexpected results based on our analysis of the idealized parent band with uniform quantum geometry.
We now argue that these lessons learned from this model can provide a unified perspective for understanding realistic systems described by the parent band paradigm.

The two important aspects of real bands that must be considered are non-uniformity of the quantum metric and Berry curvature.
We first comment on quantum metric effects.
Our results concerning topology are only based on the complex phases of the parent form factors $\arg(\mathcal{F})$ (which reflects the Berry curvature), whereas the magnitude $|\mathcal{F}|^2$ (which reflects the quantum metric) enters only as an effective rescaling of the potential or interaction strength.
Even with arbitrary $|\mathcal{F}|^2$, as long as all phases are gauge-equivalent to the uniform case of Eq~\ref{eq:Funiform}, the $\hat{W}$ and $\hat{Q}$ mappings can still be applied.
Of course, the quantum metric could affect the energetic competition between potential crystalline or liquid states~\cite{abouelkomsan2023quantum}, which we do not address here.

Since small deviations away from these ideal conditions will not close the topological gap, we expect our results regarding Chern numbers to be robust to small non-uniformities of quantum geometry.
In a real system, as long as the Berry curvature is approximately uniform near the band edge, our results should apply as long as the physics involves mostly those low-energy states.
If the Berry curvature distribution is highly non-uniform even at the band edge, the direct applicability of our theory becomes difficult to justify (indeed, counterexamples exist where the potential-induced Chern number is of the same sign as the parent Berry curvature~\cite{sheng2024quantum}).
Nevertheless, we find that the idealized model can still be surprisingly effective in capturing the qualitative physics of realistic parent bands with non-uniform Berry curvature.

% Namely, our main results are that, for a parent band with uniform Berry curvature $\mathcal{B}=2\pi N/\Omega_{\mathrm{BZ}}$, (i) a periodic potential results in a $C=-N$ Chern insulator, and (ii) strong interactions results in a $C=N$ AHC state with ideal quantum geometry.

% demonstrate that  experimental observation of the integer and fractional QAH effects in rhombohedral multilayer graphene

% We focus in particular on rhombohedral graphene multilayers to illustrate our point.
% We focus on rhombohedral multilayer graphene, 
To illustrate this point, we examine rhombohedral multilayer graphene through the lens of our theory.
 We focus on rhombohedral pentalayer graphene (R5G), with the goal of understanding the recent experimental observation of integer and fractional QAH effects in hBN-aligned R5G~\cite{lu2024fractional}.
In Fig~\ref{fig:pentalayer}a, we show the microscopic band structure of one spin-valley component of rhombohedral pentalayer graphene (R5G), expanded near the $K$ point, for several values of displacement field which is modeled as a layer potential $u_D$.
Full details of the microscopic model are available in the supplemental material~\cite{supplement}.
We remark that there is a limit in which a slightly modified microscopic wavefunction of rhombohedral multilayer graphene reduces to that the ideal parent band in Eq~\ref{eq:icbpsi}~\cite{supplement}.
In a displacement field, the conduction band edge carries a total Berry flux of $5\pi$, which has a (quite non-uniform) ring-like distribution in momentum space, as shown in Fig~\ref{fig:pentalayer}b.
We focus only on a single spin-valley polarized sector, which has positive parent Berry curvature $\mathcal{B}(\bm{k})>0$. 
We consider imposing the BZ illustrated in Fig~\ref{fig:pentalayer}b (corresponding to the experimentally relevant periodicity $a\approx \SI{11}{nm}$~\cite{lu2024fractional}).

Despite the fact that the Berry curvature is far from uniform, we now demonstrate that the properties of this parent band in the presence of interactions or a periodic potential can be qualitatively explained by our theory.
First, we may make a crude estimate of what Chern numbers to expect.
We notice that the kinetic energy increases sharply for $|\bm{k}|$ beyond $\bar{k}\approx \SI{0.5}{nm^{-1}}$, providing a natural momentum cutoff.
This gives an average Berry curvature seen by low-energy electrons of $\bar{\mathcal{B}}\approx 5\pi/(\pi\bar{k}^2)$.  
With the BZ area $\Omega_{\mathrm{BZ}}=8\pi^2/(\sqrt{3}a^2)$, this leads to an average Berry flux per BZ of $\bar{\mathcal{B}}\Omega_{\mathrm{BZ}}\approx 1.2\times 2\pi$, which is close to $2\pi$.
Based on our theory, we therefore expect a $C=\pm 1$ state at $\nu=1$ in the presence of interactions or a periodic potential, respectively.

We perform self-consistent HF calculations with gate-screened Coulomb interactions~\cite{supplement}.
The result is a gapped AHC with $C=1$ consistent with other theoretical studies~\cite{dong2023theory,zhou2023fractional,dong2023anomalous,guo2023theory,kwan2023moir}.
The sign of the Chern number is positive, in agreement with our analysis of the Fock-driven AHC.
Fig~\ref{fig:pentalayer}c shows the quantum geometric properties of the HF band as a function of the relative dielectric constant $\epsilon_r$.  
With increasing interaction strength, we find decreasing TCV.  
Note that although the emergence of the exactly ideal $\mathrm{TCV}\rightarrow 0$ Chern band is a special feature of the ideal parent band, the tendency for TCV to decrease with interaction strength is more general and can be understood as a natural consequence of exchange interaction.
To see this, we write the total Fock energy of the filled band as
\begin{equation}
\langle \mathcal{H}_{\mathrm{F}}\rangle = 
-\frac{1}{2A}\sum_{\bm{q}}V(\bm{q}) \sum_{\bm{k}\in\mathrm{BZ}}|F(\bm{k}+\bm{q},\bm{k})|^2
\end{equation}
where $F(\bm{k}+\bm{q},\bm{k})=\langle {\psi_{\lceil \bm{k}+\bm{q}\rceil}}|e^{i\bm{q}\cdot\bm{r}}\ket{\psi_{\bm{k}}}$ is the form factor of the filled band of states $\ket{\psi_{\bm{k}}}$, and $\lceil \cdot \rceil$ denotes folding back to the first BZ.  
For small $\bm{q}$, where $V(\bm{q})$ is peaked,  $|F(\bm{k}+\bm{q},\bm{k})|^2\approx 1-\sum_{\mu\nu}q_\mu q_\nu g_{\mu\nu}^{\mathrm{FS}}(\bm{k})$.
Thus, the Fock energy is optimized by minimizing the trace of the quantum metric~\cite{abouelkomsan2023quantum}, hence minimizing TCV.

Next, we consider the effect of a periodic scalar potential instead.  
We consider a honeycomb potential ($\phi=\pi$) which, based on our theory, we expect to give the most stable topological state for $\bar{\mathcal{B}}\approx2\pi/\Omega_{\mathrm{BZ}}$.
The Chern number of the first band is shown in Fig~\ref{fig:pentalayer}d as a function of potential strength $U_0$ and displacement field $u_D$.
Over a large region of the phase diagram, we find a $C=-1$ phase despite the positive parent Berry curvature, a counterintuitive result that is correctly explained by our theory.
For small $u_D$, the distribution of the Berry parent curvature becomes highly concentrated near the origin and we find regions with $C=0,1$ (indicating a transition to a state that is no longer adiabatically connected to the uniform limit).
% In this case, we find a $C=-1$ band, a counterintuitive result that is correctly explained by our theory.
% Fig~\ref{fig:pentalayer}d shows the quantum geometric properties as a function of potential strength $U_0$.  
% {\color{blue}(maybe change part d to show phase diagram)}
% Furthermore, this $C=-1$ band persists over a wide range of $u_D$ and potential strength and shape~\cite{supplement}, so this result is not fine-tuned.  
For the $u_D$ relevant to the AHC, however, the $C=-1$ state is quite robustly favored by the periodic potential.
A strong periodic potential will therefore lead to a topologically distinct state than the AHC in R5G.

% These states were only observed for one sign of displacement field
% and theoretical works following~\cite{ lu2024fractional,dong2023theory,zhou2023fractional,dong2023anomalous,guo2023theory,kwan2023moir}.
% In this system, a moir\'e superlattice is present only on one side, while the topological states were only observed
 % when the displacement field localizes the electrons to the opposite side, furthest from the moir\'e.

% We now comment on the observation of integer and fractional QAH states in R5G/BN~\cite{lu2024fractional}.
Our theory provides a unified perspective to understand several puzzling aspects of the experiment on R5G/hBN~\cite{lu2024fractional}.
The observed $C=1$ QAH state is most naturally viewed as an AHC that is weakly pinned by the moir\'e.  
Although the moir\'e potential from hBN also contains a sublattice-dependent term, the most significant portion of it is simply a periodic scalar potential $U(r)$~\cite{supplement} (other contributions include out-of-plane polarization due to charge transfer at the graphene-hBN interface, and Hartree corrections due to the background induced charge density, both of which also contribute as an effective scalar potential).
While a weak pinning potential should help stabilize the state, our theory predicts that a strong periodic potential competes with the AHC and eventually destabilizes it.
This provides an explanation to the puzzle of why the QAH state is observed only when electrons are localized by the displacement field to the layer furthest from the moir\'e interface~\cite{lu2024fractional} (where the moir\'e potential is weakest).  
In fact, in this case there is a sharp distinction between the exchange-driven AHC and a possible (more conventional) moir\'e-driven Chern insulator~\cite{Parameswaran2024comment}, given by the relative sign of the miniband Chern number and parent Berry curvature.
%: for the opposite sign of displacement field, and the effective periodic potential is sufficiently strong to destabilize the QAH state.
We also comment on the physics at fractional filling.
Exchange interactions naturally favor minimizing the quantum metric of the AHC (and in our idealized model, lead to the formation of a perfect Landau-level-like band).
This natural tendency leads to bands that are favorable for realizing fractional QAH states at partial filling, as observed experimentally.

% First, the QAH states were only observed in a strong displacement field that localizes the parent band to the layer furthest from the graphene-hBN moir\'e interface, but are absent for the opposite sign of displacement field.  
% Thus, a strong moir\'e potential appears to be detrimental to the QAH state, which is puzzling as one might expect the AHC to be.  

% Distinguish between Chern insulator and AHC.

% Fractional: coexistence of QHFM and 

% Naively, one would have expected that the topological states should 

% Our theory gives an explanation for why the integer and fractional QAH states were observed only for one sign of displacement field~\cite{lu2024fractional}, which is counterintuitively only when the parent band is located on the layer farthest from the moir\'e superlattice.

}

\section{Outlook}
There are many interesting directions for further study of our model, as well as possible extensions.
Although we have focused in this paper on the case where either the potential or interactions dominate, 
there may be intermediate phases between the two that would not have an analog in the trivial parent band model.
Another interesting question is the role of parent Berry curvature distribution:
while we have focused only on the uniform case, the effect of non-uniform Berry curvature can be systematically studied by exploring the finite-$S$ models and tuning $M$.
It would also be interesting to consider a ``valley-ful'' extension of our model, in which we allow for a time-reversed partner of the parent band, which could
 allow for interesting new states or excitations.

The exact $\mathcal{B}\rightarrow\mathcal{B}+2\pi/\Omega_{\mathrm{BZ}}$ periodicity of the many-body interacting Hamiltonian $\hat{\mathcal{H}}=\hat{\mathcal{H}}_0+\hat{\mathcal{H}}_{\mathrm{F}}$
implies that all properties of the AHC in this model can be obtained from the properties of the corresponding WC.
This includes the properties of all excited states as well.
It would be interesting to analyze the resulting dynamics ~\cite{Gruner1988dynamics} of the AHC through the lens of this mapping.
To this end, it is therefore important to determine the effects of the interaction terms that were neglected in making the Fock-dominated $\hat{\mathcal{H}}_{\mathrm{int}}\approx\mathcal{H}_{\mathrm{F}}$ approximation. 
While our numerical HF analysis of the ground state with Coulomb potential is in good agreement with the physics being Fock-dominated, the neglected terms are likely to be important for other physics, such as those of excitations above the ground state.

{ The parent band perspective presented in this work may also be useful for engineering future topological quantum phases.
For example, we anticipate that many new topological states can be realized by combining the high parent Berry curvature of rhombohedral multilayer graphene with engineered electrostatic superlattice potentials, as has been realized with a wide variety of experimental techniques~\cite{forsythe2018band,shi2019gate,xu2021creation,barcons2022engineering,sun2023signature,li2021anisotropic,song2015topological,su2022massive,ghorashi2023topological,ghorashi2023multilayer,zeng2024gatetunable,tan2024designing,gu2023remote,zhang2024engineering,he2024dynamically,kim2023electrostatic,yasuda2021stacking,wang2022interfacial,zhao2021universal,woods2021charge,wang2024band,zhang2024emergence,ding2024engineering}.

% We may ask what the implications of band geometry of the AHC is, in the absence of a pinning potential.
Of particular importance is understanding the fractionalized phases that could result from the AHC.
While the HF band geometry may be favorable for fractionalized states, the concept of fractional filling becomes somewhat dubious in the absence of a pinning potential, since the unit cell is spontaneously determined.
Nevertheless, an interesting observation is that the Berry curvature density $\mathcal{B}$ itself sets a natural unit cell area, $\Omega_{\mathrm{BZ}}=2\pi/\mathcal{B}$, in which there is $2\pi$ Berry flux per BZ.  
Now, consider filling electrons to density corresponding to fractional filling $p/q<1$ of $\Omega_{\mathrm{BZ}}$.
This density can also be interpreted as full filling of an alternate unit cell $\Omega_{\mathrm{BZ}}^\prime=(p/q)\Omega_{\mathrm{BZ}}$ with an effective Berry flux of $\mathcal{B}\Omega_{\mathrm{BZ}}^\prime=(p/q)2\pi$ per BZ.
 There are therefore (at least) two distinct ways for translation symmetry to be broken: (1) with the $\Omega_{\mathrm{BZ}}$ periodicity, which could result in a partially filled AHC band, or (2) with the $\Omega_{\mathrm{BZ}}^\prime$ periodicity, which could lead to a WC or AHC.
Which of these is realized will depend on energetic details.
% we speculate that the system may prefer fractionally filling the $\Omega_{\mathrm{BZ}}$ unit cell, rather than full filling of the $\Omega_{\mathrm{BZ}}^\prime$ unit cell. 
If the first scenario is preferred, the partially filled ideal AHC band naturally leads to the possibility of a ``fractional anomalous Hall crystal'' that spontaneously breaks continuous translation symmetry to form a fractional Chern insulator (the continuum limit of previously studied lattice phases~\cite{kourtis2014combined,kourtis2018symmetry}).
The ideal parent band introduced in this work would be well suited for the theoretical study of this exotic state.
% {\color{blue}QHFM picture.}

% \subsection{Conclusion}

We conclude on a philosophical note.
The modern theoretical understanding of correlated physics in flat Chern bands is largely built on foundational breakthroughs in the quantum Hall effect.
% We stand on the shoulders of giants.
We would not try to understand fractional Chern insulators before understanding the fractional quantum Hall effect in Landau levels.
In this work, our primary motivation has been that the non-interacting starting point for the fractional Chern insulators observed in rhombohedral graphene is fundamentally unlike the Landau level.
Instead, the physical phenomenon at play appears to be that of anomalous Hall crystallization.
Rather than trying to understand the specific material system at hand, our approach has been to first understand the analogous ``Landau level'' of this phenomenon: the ideal parent band.
Just as the rich structure of Landau levels laid the foundation for the theory of the quantum Hall effect,
we believe that the ideal parent band can provide the groundwork for a comprehensive theoretical understanding of these new topological phases of matter.
% The rich structure of Landau levels led to the foundation of topological phases.
% We anticipate that the similarly rich structure of the ideal parent band has the potential to lead to a the theoretical foundation for the topological phases.
}

% {\color{blue} Engineering future devices.}

% {\color{blue} Mechanism for topological state in R5G is entirely new.  
% Studying the ICB is like studying the Landau level, we wouldn't try to understand Chern insulators before understanding the QH effect.}

\emph{Note added ---}
While finalizing our manuscript, and shortly afterwards, several related preprints appeared~\cite{zeng2024sublattice,soejima2024anomalous,dong2024stability} also studying the formation of the AHC;
while there is little direct overlap, our findings are all consistent.

\acknowledgements
We are indebt to Liang Fu for valuable discussions at the inception of this project.
We acknowledge helpful discussions with Aidan P. Reddy, Yves Kwan, 
Sid Parameswaran, Steve Kivelson, Sri Raghu, Senthil Todadri, Mike Zaletel, Dan Parker, Julian May-Mann, Patrick Ledwith, Junkai Dong, Zi-ang Hu, and Tomohiro Soejima. TD acknowledges support from a startup fund at Stanford University. TT is supported by the Stanford Graduate Fellowship. The computations are performed using resources provided by the Stanford Research Computing Center.

\bibliography{ref}

\appendix

\onecolumngrid

\section{ Trace condition violation of ansatz wave functions}
We showed numerically in the main text figure that with the gauge transformation, $\hat{Q}$ (Fock) and $\hat{W}$ (Potential), the TCV of the $C=\pm1$ band decrease/increase as a function of $\xi$ at large $\xi$. We now derive this analytically. Trace condition violation is defined as following\par

\begin{equation}
\mathrm{TCV}\equiv\frac{1}{2\pi}\int (\Tr[g_{\mu\nu}^{\mathrm{FS}}(\bm{k})]-|B(\bm{k})|)d \bm{k}
\end{equation}
We first study the behaviour of $\int \Tr[g_{\mu\nu}^{\mathrm{FS}}(\bm{k})]d \bm{k}$ for two ansatz wave functions when $\xi$ is large.  We focus on the square lattice case to avoid cluttered notation, though the proof can be easily generalized to triangular lattice.

The properly normalized $\hat{Q}$ ansatz wavefunction $\ket{\psi^{-}(\bm{k})}$ is
\begin{equation}
\ket{\psi^{-}(\bm{k})}
= \sum_{\bm{g}}\mathcal{N}(\bm{k})e^{-\frac{(\bm{k}+\bm{g})^2}{4\xi^2}}e^{-i\frac{\pi}{g^2}[g_xg_y+k_xg_y-k_yg_x]}\ket{u_{\bm{k}+\bm{g}}}e^{i\bm{g}\cdot\bm{r}}
\end{equation}
with normalization constant
\begin{equation}
\mathcal{N}(\bm{k})=\frac{g}{\sqrt{2\pi}\xi}[\mathcal{\theta}(z_x,q)\mathcal{\theta}(z_y,q)]^{-\frac{1}{2}}
\end{equation}
where  $z_{x,y}\equiv \frac{k_{x,y}\pi}{g}$, $q\equiv \exp(-\frac{2\pi^2\xi^2}{g^2})$, and $\theta$ is the Jacobi theta function $\theta(z,q)=\sum_{n=-\infty}^{\infty}q^{n^2}\exp(2inz)$

Notice that $|\braket{\psi^{-}(\bm{k})}{\psi^{-}(\bm{k}+\bm{\delta})}|^2=1-g_{\mu\nu}\delta_\mu \delta_{\nu}$, where $\bm{\delta}$ is an infinitesimally small vector, and we have kept only the leading order terms. 

First, we examine the $g_{xx}$ term.  Taking $\bm{\delta}=(\delta_x,0)$, we have

\begin{equation}
\begin{split}
|\braket{\psi^{-}(\bm{k})}{\psi^{-}(\bm{k}+(\delta_x,0))}|^2
&=\left|\sum_{\bm{g}}\frac{1}{\mathcal{N}(\bm{k})\mathcal{N}(\bm{k}+(\delta_x,0))}e^{-\frac{(\bm{k}+\bm{g})^2}{2\xi^2}}e^\frac{{-\delta_x^2}}{4\xi^2}e^{-\frac{(k_x+g_x)\delta_x}{2\xi^2}}e^{-i\frac{\pi}{g^2}\delta_x g_y}e^\frac{-\pi \delta_x^2}{2g^2}
e^{\frac{i\pi (k_y+g_y)\delta_x}{2g^2}}\right|^2\\
&=e^{\frac{-\pi^2 (\delta^2_x)}{g^2}}e^{-\frac{\delta^2_x}{2\xi^2}}\left|\sum_{\bm{g}}\frac{1}{\mathcal{N}(\bm{k})\mathcal{N}(\bm{k}+(\delta_x,0))}e^{-\frac{(\bm{k}+\bm{g})^2}{2\xi^2}}e^{-\frac{(k_x+g_x)\delta_x}{2\xi^2}}\right|^2\\
&=e^{\frac{-\pi^2 (\delta^2_x)}{g^2}}e^{-\frac{\delta^2_x}{2\xi^2}}
\frac{e^{\frac{\delta^2_x}{4\xi^2}}\theta(z_x+\frac{\delta_x\pi}{2g},q)\theta(z_y,q)^2}{\theta(z_x,q)\theta^2(z_y,q)\theta(z_x+\frac{\delta_x \pi}{g},q)}\\
&=1-\frac{\pi^2\delta_x^2}{g^2}-\frac{\delta^2_x}{4\xi^2}+\frac{2\pi^2\delta^2_x}{g^2}q\cos(\frac{2\pi k_x}{g^2})+O(\delta^3,q^2)
\end{split}
\end{equation}

\begin{equation}
g_{xx}(\bm{k})=\frac{\pi^2}{g^2}+\frac{1}{4\xi^2}-2\pi^2 \frac{q}{g^2}cos(2\pi \frac{k_x}{g^2})+O(q^2)
\end{equation}
Taking $\bm{\delta}$ to be along y direction will give similar expression for $g_{yy}$. Then,

\begin{equation}
\Tr[g^{\mathrm{FS}}_{\mu\nu}(\hat{Q})]=\frac{2\pi}{g^2}+\frac{1}{2\xi^2}
-\frac{2\pi^2q}{g^2}\cos\left(\frac{2k_x\pi}{g}\right)-\frac{2\pi^2q}{g^2}\cos\left(\frac{2k_y\pi}{g}\right)+O(q^2)
\end{equation}
At large $\xi$, all the momentum dependence are suppressed by $q=\exp(-\frac{2\pi^2\xi^2}{g^2})$, which is negligible compared with $\frac{1}{\xi^2}$.

For the $\hat{W}$ ansatz, a similar analysis reveals 
\begin{equation}
\begin{split}
\Tr[g^{\mathrm{FS}}_{\mu\nu}(\hat{W})]&=\frac{2\pi}{g^2}+\frac{1}{2\xi^2}+\frac{8\pi^2\xi^2}{g^4}-\frac{2\pi^2q}{g^2}\cos\left(\frac{2k_x \pi}{g}\right)-\frac{2\pi^2q}{g^2}\cos\left(\frac{2k_y\pi}{g}\right)\\
&-\frac{32\pi^4\xi^4q}{g^6}\cos\left(\frac{2k_x\pi}{g}\right)-\frac{32\pi^4\xi^4q}{g^6}\cos\left(\frac{2k_y\pi}{g}\right)+O(q^2)
\end{split}
\end{equation}
Again, all the momentum dependence is suppressed by powers of $q$, and are therefore exponentially suppressed in $\xi$.\par

We now study the second term $\int d\bm{k}|B(\bm{k})|$ in the definition of TCV, and prove that momentum dependence of $|B(\bm{k})|$ is exponentially suppressed by $\xi$. To leading orders in $q$ and $\delta_x$
\begin{equation}
\braket{\psi^{+}(\bm{k}+(\delta_x,0))}{\psi^{+}(\bm{k})}=1+i\delta_x\frac{\pi k_y}{g^2}-i 8\pi \xi^2 \frac{q}{g^3}\delta_x \sin(2\pi \frac{k_y}{g})+O(q^2)
\end{equation}
Then the corresponding connection is
\begin{equation}
\alpha^{+}_x(\bm{k})=-i\lim_{\delta_x\to0}
\frac{\braket{\psi^{+}(\bm{k}+(\delta_x,0))}{\psi^{+}(\bm{k})}-1}{\delta_x}=\pi \frac{k_y}{g^2}+O(q)
\end{equation}
By symmetry, the leading order momentum-independent Berry curvature is
\begin{equation}
B^{+}(\bm{k})=-2\partial_{k_y}\alpha_x(\bm{k})=-\frac{2\pi}{g^2}+O(q)\quad\quad 
C^{+}=\frac{\int_{BZ} d\bm{k}B^{+}(\bm{k})}{2\pi}=-1
\end{equation}
All  momentum dependence of $B^{+}(\bm{k})$ are suppressed by powers of $q$, and are there for negligible at large $\xi$. The same is also true for $B^{-}(\bm{k})$. Then for both ansatz at large $\xi$,
\begin{equation}
\int d\bm{k}|B^{\pm}(\bm{k})|=2\pi
\end{equation}

\section{Change of Chern number under $\hat{W}$ and $\hat{Q}$}

In the main text, we sketched how the transformation $\hat{W}$ and $\hat{Q}$ will change the Chern number and confirmed it numerically. In this section, we show this analytically. We focus on the strong-potential ($\hat{W}$), square lattice case, though the results can be straightforwardly extended to other cases.

The Hamiltonian at $\mathcal{B}_0$ with potential strength $U_0$ takes the form, $\hat{\mathcal{H}}=\hat{\mathcal{H}}_0+\hat{\mathcal{H}}_{\mathrm{pot0}}$.
\begin{equation}
\begin{split}
\hat{\mathcal{H}}_0=&\sum_{\bm{k},\bm{g}}\mathcal{E}(\bm{k}+\bm{g})c^{\dagger[\mathcal{B}_0]}_{\bm{k},\bm{g}}c^{[\mathcal{B}_0]}_{\bm{k},\bm{g}}\\
\hat{\mathcal{H}}_{\mathrm{pot0}}=&-U_0e^{-\frac{\mathcal{B}_0}{4}g^2}\sum_{\bm{k},\bm{g},j} c^{\dagger[\mathcal{B}_0]}_{\bm{k},\bm{g}+\bm{b}_j} e^{i\frac{\mathcal{B}_0}{2}(\bm{k}+\bm{g})\cross\bm{b}_j}c^{[\mathcal{B}_0]}_{\bm{k},\bm{g}}
\end{split}
\end{equation}
This Hamiltonian is solved by diagonalizing the matrix $H_{\bm{g}^\prime\bm{g}}(\bm{k})$ at each $\bm{k}$, 
\begin{equation}
\hat{\mathcal{H}}=\sum_{\bm{k}\bm{g}\bm{g'}}c^{\dagger[\mathcal{B}_0]}_{\bm{k},\bm{g'}}H_{\bm{g'}\bm{g}}(\bm{k})c^{[\mathcal{B}_0]}_{\bm{k},\bm{g}}
\end{equation}

\begin{equation}
H_{\bm{g'}\bm{g}}(\bm{k})=\mathcal{E}(\bm{k}+\bm{g})\delta_{\bm{g'}\bm{g}}-\sum_{\bm{b}_j}U_0e^{-\frac{\mathcal{B}_0}{4}g^2}e^{i\frac{\mathcal{B}_0}{2}(\bm{k}+\bm{g})\cross\bm{b}_j}\delta_{\bm{g'}-\bm{g},\bm{b}_j}
\end{equation}
The Hamiltonian at $\mathcal{B}_1=\mathcal{B}_0+\frac{2\pi}{g^2}$ with potential strength $U_0\exp(\frac{\pi}{2})$ takes the form, $\hat{\mathcal{H}_1}=\hat{\mathcal{H}}_{01}+\hat{\mathcal{H}}_{\mathrm{pot1}}$, with
\begin{equation}
\begin{split}
\hat{\mathcal{H}}_{01}=&\sum_{\bm{k},\bm{g}}\mathcal{E}(\bm{k}+\bm{g})c^{\dagger[\mathcal{B}_1]}_{\bm{k},\bm{g}}c^{[\mathcal{B}_1]}_{\bm{k},\bm{g}}\\
\hat{\mathcal{H}}_{\mathrm{pot1}}=&-U_0e^{-\frac{\mathcal{B}_0}{4}g^2}\sum_{\bm{k},\bm{g},j} c^{\dagger[\mathcal{B}_1]}_{\bm{k},\bm{g}+\bm{b}_j} e^{i\frac{\mathcal{B}_1}{2}(\bm{k}+\bm{g})\cross\bm{b}_j}c^{[\mathcal{B}_1]}_{\bm{k},\bm{g}}
\end{split}
\end{equation}
 where we stress that $c^{\dagger[\mathcal{B}]}_{\bm{k},\bm{g}}$ creates $e^{i\bm{k}\cdot \bm{r}}\ket{s^{\mathcal{B}}_{\bm{k+g}}}$. We define
\begin{equation}
d^{\dagger[\mathcal{B}_1]}_{\bm{k},\bm{g}}\equiv c^{\dagger[\mathcal{B}_1]}_{\bm{k},\bm{g}} W_{\bm{g}}(\bm{k})
\end{equation}
The definition of the phase $W_{\bm{g}}(\bm{k})$  is given in the main text:
 
 \begin{equation}
W_{\bm{g}}(\bm{k})=\exp{i\pi\left(\frac{\bm{k}\cross\bm{g}}{\Omega_{\mathrm{BZ}}}+\omega(\bm{g})\right)} \quad \omega(\bm{g})=\frac{g_xg_y}{\Omega_{\mathrm{BZ}}},
\end{equation}
$\hat{\mathcal{H}_1}$, written in terms of operators $d^{\dagger}$ and $d$ operators, is formally identical to  $\hat{\mathcal{H}}$, written in terms of operators $c^{\dagger[\mathcal{B_0}]}$ and $c^{[\mathcal{B_0}]}$. Specifically, the matrix we need to diagonalize when solving this non-interacting Hamiltonian is exactly the same in these two cases, and hence the  eigenvalues and  eigenvectors are identical in $\mathcal{B}_0$ and $\mathcal{B}_1$ case. To see this point, just consider one example term when $\bm{g}=n_1\bm{b}_1+n_2\bm{b}_2$ \par
\begin{equation}
\begin{split}
&c^{\dagger[\mathcal{B}_1]}_{\bm{k},\bm{g}+\bm{b}_2} e^{i\frac{\mathcal{B}_1}{2}(\bm{k}+\bm{g})\cross\bm{b}_2}c^{[\mathcal{B}_1]}_{\bm{k},\bm{g}}\\
&=c^{\dagger[\mathcal{B}_1]}_{\bm{k},\bm{g}+\bm{b}_2} \exp(i\frac{\mathcal{B}_0}{2}(\bm{k}+\bm{g})\cross\bm{b}_2)
\exp(i\pi(\frac{\bm{k}\times(\bm{g}+\bm{b}_2)}{\Omega_{BZ}}+n_1(n_2+1)))
\exp(-i\pi(\frac{\bm{k}\times\bm{g}}{\Omega_{BZ}}+n_1n_2))
c^{[\mathcal{B}_1]}_{\bm{k},\bm{g}}\\
&=d^{\dagger[\mathcal{B}_1]}_{\bm{k},\bm{g}+\bm{b}_2} e^{i\frac{\mathcal{B}_0}{2}(\bm{k}+\bm{g})\cross\bm{b}_2}d^{[\mathcal{B}_1]}_{\bm{k},\bm{g}}
\end{split}
\end{equation}
Notice, crucially, this equality depends on the fact that we keep only the the first harmonics in the periodic potential.\par

 It should be noted that the eigenvectors in these two cases are written in two different basis. If the Bloch wavefunctions at $\mathcal{B}_0$ for a single isolated band are
\begin{equation}
|\psi_{\bm{k}}^{(0)}\rangle=\sum_{\bm{g}}\psi_{\bm{g}}^{(0)}(\bm{k})e^{i(\bm{k}+\bm{g})\cdot \bm{r}}|s^{\mathcal{B}_0}_{\bm{k}+\bm{g}}\rangle 
\end{equation}
where  $\psi^{(0)}_{\bm{g}}(\bm{k})$ is the eigenvector of the matrix $H_{\bm{g}^\prime\bm{g}}(\bm{k})$.
% while the physical wavefunction is written in the basis of $\ket{u_{\bm{k+g}}^{\mathcal{B}_0}}$.
 By the mapping we discussed above, the corresponding wavefunction at $\mathcal{B}_1$ is
\begin{equation}
|\psi_{\bm{k}}^{(1)}\rangle=\sum_{\bm{g}}\psi_{\bm{g}}^{(0)}(\bm{k})e^{i(\bm{k}+\bm{g})\cdot \bm{r}}W_{\bm{g}}(\bm{k})|s^{\mathcal{B}_1}_{\bm{k}+\bm{g}}\rangle 
\end{equation}

The Berry curvature of $|\psi_{\bm{k}}^{(0)}\rangle$ is 
\begin{equation}
\begin{split}
B^{(0)}(\bm{k})&=i\epsilon_{\mu\nu}[\partial_\mu(\langle{\psi^{(0)}_{\bm{k}}}|e^{i\bm{k}\cdot\bm{r}})][\partial_\nu(e^{-i\bm{k}\cdot\bm{r}}|{\psi^{(0)}_{\bm{k}}}\rangle)]\\
&=b_{01}(\bm{k})+b_{02}(\bm{k})+b_{03}(\bm{k})
\end{split}
\end{equation}
where $\mu$ takes values $x$ and $y$
\begin{equation}
\begin{split}
&b_{01}(\bm{k})=i\epsilon_{\mu\nu}[\partial_{\mu}\psi^{(0)*}_{\bm{g}}(\bm{k})][\partial_{\nu}\psi^{(0)}_{g}(\bm{k})]\\
&b_{02}(\bm{k})=\sum_{\bm{g}}\epsilon_{\mu\nu}[\partial_\mu |\psi^{(0)}_{\bm{g}}(\bm{k})|^2] A_\nu(\bm{k}+\bm{g})\\
&b_{03}(\bm{k})=\sum_{\bm{g}}|\psi^{(0)}_{\bm{g}}(\bm{k})|^2 F(\bm{k}+\bm{g})\\
\end{split}
\end{equation}
and we have defined the following two parent band quantities
\begin{equation}
\begin{split}
&A_{\mu}(\bm{k})=i\braket{s^{\mathcal{B}_0}_{\bm{k}}}{\partial_{\mu} s^{\mathcal{B}_0}_{\bm{k}}}\\
&F(\bm{k})=\epsilon_{\mu\nu}\partial_\mu A_\nu(\bm{k})
\end{split}
\end{equation}

 Notice that the above three sets of equations are completely general, even if the parent band spinors $\ket{s^{\mathcal{B}_0}_{k+g}}$ are replaced by some other spinors. 
 However, $b_{02}$ and $b_{03}$ will not contribute to the total Chern number, as their sum is  a total derivative of a periodic function (using $|\psi^{(0)}_{\bm{g}+\bm{g_i}}(\bm{k})|^2=|\psi^{(0)}_{\bm{g}}(\bm{k}+\bm{g_i})|^2$)
 \begin{equation}
 b_{02}(\bm{k})+b_{03}(\bm{k})=\sum_{\bm{g}}\epsilon_{\mu\nu}\partial_\mu [|\psi^{(0)}_{\bm{g}}(\bm{k})|^2 A_\nu(\bm{k}+\bm{g})]
 \end{equation}
The term on the right is a total derivative of a periodic function
 
  \begin{equation}
 \begin{split}
 &\sum_{\bm{g}}[|\psi^{(0)}_{\bm{g}}(\bm{k}+\bm{g}_i)|^2 A_\nu(\bm{k}+\bm{g}+\bm{g}_i)]= \sum_{\bm{g}}[|\psi^{(0)}_{\bm{g}}(\bm{k})|^2 A_\nu(\bm{k}+\bm{g})]\\
 &\Rightarrow \int d\bm{k}[b_{02}(\bm{k})+b_{03}(\bm{k})]=0
 \end{split}
 \end{equation}
 
Thus, we arrive at the following conclusion
\begin{equation}
2\pi C^{(0)}=\int_{BZ} B^{(0)} (\bm{k}) d\bm{k}=\int_{BZ} b_{01} (\bm{k}) d\bm{k}
\end{equation}

Correspondingly for $\ket{\psi^{(1)}_{\bm{k}}}$
\begin{equation}
\begin{split}
2\pi C^{(1)}
=&\int_{BZ}  \sum_{\bm{g}}i\epsilon_{\mu\nu}[\partial_{\mu}(W^{*}_{\bm{g}}(\bm{k})\psi^{(0)*}_{\bm{g}}(\bm{k}))][\partial_{\nu}(W_{\bm{g}}(\bm{k})\psi^{(0)}_{g}(\bm{k}))]d\bm{k}\\
=&\int_{BZ}  \sum_{\bm{g}}i\epsilon_{\mu\nu}\partial_{\mu}\Biggl\{W^{*}_{\bm{g}}(\bm{k})\psi^{(0)*}_{\bm{g}}(\bm{k})[\partial_{\nu}(W_{\bm{g}}(\bm{k})\psi^{(0)}_{g}(\bm{k}))]\Biggl\}d\bm{k}\\
=&\int_{BZ} \sum_{\bm{g}} i\epsilon_{\mu\nu}\partial_{\mu} (\psi^{(0)*}_{\bm{g}}(\bm{k}) \partial_{\nu}\psi^{(0)}_{\bm{g}}(\bm{k}) )d\bm{k}\\
&+\sum_{\bm{g}} i\epsilon_{\mu\nu}\partial_{\mu}[|\psi^{(0)}_{\bm{g}}(\bm{k})|^2 (W^{*}_{\bm{g}}(\bm{k}) \partial_{\nu}W_{\bm{g}}(\bm{k}) )]d\bm{k}\\
=&2\pi C^{(0)}+\int_{BZ}\sum_{\bm{g}} i\epsilon_{\mu\nu}\partial_{\mu}[|\psi^{(0)}_{\bm{g}}(\bm{k})|^2 (W^{*}_{\bm{g}}(\bm{k}) \partial_{\nu}W_{\bm{g}}(\bm{k}) )]d\bm{k}
\end{split}
\end{equation}

We now focus on the second term on the last line. 
Defining
\begin{equation}
\alpha_{\nu}(\bm{k})\equiv\sum_{\bm{g}}|\psi^{(0)}_{\bm{g}}(\bm{k})|^2 (W^{*}_{\bm{g}}(\bm{k}) \partial_{\nu}W_{\bm{g}}(\bm{k}) )
\end{equation}
 the change in the Chern number is given by
\begin{equation}
2\pi C^{(1)}=2\pi C^{(0)}+\int_{0}^{g} i[\alpha_{y}(\bm{k}+g(1,0))-\alpha_{y}(\bm{k})]dk_y-\int_{0}^{g} i[\alpha_{x}(\bm{k}+g(0,1))-\alpha_{x}(\bm{k})]dk_x
\end{equation}

With the form of $W_{\bm{g}}(\bm{k})$ given in the main text, $\alpha_{\mu}(\bm{k})$ has the following behaviour under the translation of reciprocal lattice vectors
\begin{equation}
\begin{split}
\alpha_x(\bm{k}+g(0,1))=\alpha_x(\bm{k})-i\frac{\pi}{g}\\
\alpha_y(\bm{k}+g(1,0))=\alpha_y(\bm{k})+i\frac{\pi}{g}
\end{split}
\end{equation}
which leads to $C^{(1)}=C^{(0)}-1$. This finishes the proof.
% \par
% Finally, we note that these calculations (square lattice and triangular lattice), can all be done in a gauge-independent manner.

\section{Hartree-Fock calculation on ideal parent band}
We will describe the operational perspective of the Hartree-Fock calculation presented in the main text on the ideal parent band. This is different from the band-basis Hartree-Fock carried out on pentalayer graphene, which is discussed in Sec~\ref{sec:pentalayerhf}

The Hartree-Fock approximation amounts to to replacing the interaction by Hartree term and Fock term, $\hat{\mathcal{H}}_{int}\approx \hat{\mathcal{H}}_{H}+\hat{\mathcal{H}}_{F}$, defined in the main text, and performing the mean-field approximation on the four-fermion operators.
The mean-field Hamiltonian is
\begin{equation}
\begin{split}
\hat{\mathcal{H}}_{\mathrm{H}}^{\mathrm{mf}}=\frac{1}{A}\sum_{\substack{\bm{k}_1\bm{k}_2\\\bm{g}_1\bm{g}_2\bm{g}_3\bm{g}_4}} \tilde{V}^{\bm{k}_1\bm{k}_2\bm{k}_2\bm{k}_1}_{\bm{g}_1\bm{g}_2\bm{g}_3\bm{g}_4}
\mathcal{P}_{\bm{g_4},\bm{g_1}}(\bm{k_1})({c}_{\bm{k}_2\bm{g}_2}^\dagger  {c}_{\bm{k}_2\bm{g}_3})\\
\hat{\mathcal{H}}_{\mathrm{F}}^{\mathrm{mf}}=-\frac{1}{A}\sum_{\substack{\bm{k}_1 \bm{k}_2\\\bm{g}_1\bm{g}_2\bm{g}_3\bm{g}_4}} \tilde{V}^{\bm{k}_1\bm{k}_2\bm{k}_1\bm{k}_2}_{\bm{g}_1\bm{g}_2\bm{g}_3\bm{g}_4}
\mathcal{P}_{\bm{g_3},\bm{g_1}}(\bm{k_1})({c}_{\bm{k}_2\bm{g}_2}^\dagger  {c}_{\bm{k}_2\bm{g}_4})
\end{split}
\end{equation}
The total mean-field Hamiltonian is then $\hat{\mathcal{H}}_{HF}=\hat{\mathcal{H}}_0+\hat{\mathcal{H}}_{pot}+\hat{\mathcal{H}}_{H}^{\mathrm{mf}}+\hat{\mathcal{H}}_{F}^{\mathrm{mf}}$
We have defined the projector
\begin{equation}
\mathcal{P}_{\bm{g}_4,\bm{g}_1}(\bm{k}_1) = \langle c^{\dagger}_{\bm{k_1g_1}}c_{\bm{k_1g_4}}
\rangle
\end{equation}
which is to be evaluated in the Slater determinant ground state of $\hat{\mathcal{H}}_{\mathrm{HF}}$.

 % $\psi^n_{g}(\bm{k})$ are the eigenvectors obtained by diagonalizing $\hat{\mathcal{H}}_{HF}$ in the $c^{\dagger}_{\bm{k},\bm{g}}$, $c_{\bm{k},\bm{g}}$ basis, $\bm{g}$ is the reciprocal lattice (RL) vector, and $\bm{k}$ labels the momentum in the BZ. The Hartree-Fock orbitals are written in terms of the $c^{\dagger}$ basis as

% \begin{equation}
% d^{\dagger}_{\bm{k},n}=\sum_{\bm{g}}\psi^{n}_{\bm{g}}c^{\dagger}_{\bm{k,\bm{g}}}
% \end{equation}
% where n labels the band, and the Slater determinant ground state is constructed as
% \begin{equation}
% \ket{\mathrm{Slater}}=\prod_{\bm{k},n \in occ} d^{\dagger}_{\bm{k},n}\ket{\Omega}
% \end{equation}

% The  $\langle...\rangle$  in the definition of the projector is with respect to the the Slater determinant ground state, thus by self-consistency
% \begin{equation}
% \mathcal{P}_{\bm{g}_4,\bm{g}_1}(\bm{k}_1)=\langle c^{\dagger}_{\bm{k_1g_1}}c_{\bm{k_1g_4}}
% \rangle
% =\sum_{n\in occ}\psi^{n}_{\bm{g_4}}(\bm{k}_1)\psi^{*n}_{\bm{g_1}}(\bm{k}_1)
% \end{equation}

% The sum over n is over the set of occupied bands.  We  set $n=1$ electrons per unit cell for main text calculation.

Since $\hat{H}_{HF}$ and $\mathcal{P}$ depends on each other,
the self-consistet HF equations are solved iteratively until convergence is achieved, (which is determined by the convergence of the projector matrix). 
A random initial projector is used in the calculation.

We solve the HF equations on a system of $9\times9$ unit cells. 
Equivalently, this corresponds to a $9\times 9$ mesh of $\bm{k}$-points in the BZ, 
\begin{equation}
\bm{k}=\frac{n_1}{9}\bm{b}_1+\frac{n_3}{9}\bm{b}_3
\end{equation}
where $n_1,n_3=0,...,8$.
For the RL vectors $\bm{g}$, we keep all $|\bm{g}|\leq 4|\bm{b}_1|$ in the calculation. 
We remark that the real space charge density distribution $\rho(\bm{r})$, is calculated by
\begin{equation}
\rho(\bm{r})=\sum_{\bm{k},\bm{g},\bm{g'}}\mathcal{P}_{\bm{g},\bm{g}'}(\bm{k})\mathcal{F}(\bm{k}+\bm{g'},\bm{k}+\bm{g})e^{i(\bm{g}-\bm{g'})\cdot\bm{r}}
\end{equation}

\begin{figure}[t]
\begin{center}
\includegraphics[width=0.3\textwidth]{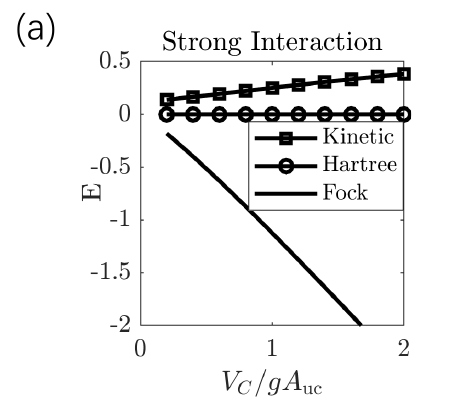}
\end{center}
\caption{Kinectic, Hartree, and Fock Energy per particle obtained from self-consistent Hartree-Fock calculation. 
The set up is the same as in main text Fig. 3. 
All energies are measured in units of $\frac{g^2}{2m}$. 
The calculation is done on a 9$\times$9 mesh.
}
\label{SPfig_HFresolvedenergy}
\end{figure}

Finally, in  Fig.~\ref{SPfig_HFresolvedenergy}, we show the contributions to the total energy of the kinetic, Hartree, and Fock terms (per particle) corresponding to the Hartree-Fock calculation performed in the main text, measured in units of $\frac{g^2}{2m}$.  
We find that the Hartree energy is orders of magnitude smaller than the kinetic and Fock terms.
% Fock energy dominates over Hartree energy in the entire range of parameters considered.
%justifying the $\hat{Q}$ transformation discussed in the main text.

\section{form factors of Landau Level}
\subsection{Preliminary}
We mentioned in the main text that the form factors of our parent band is exactly that of the lowest Landau Level. We give a concise derivation of this fact. For reference, the appendix of \cite{reddy2024nonabelian,PhysRevLett.127.246403}\footnote{The author specifically thanks A. P. Reddy for teaching him these knowledge related with Landau Level.}, contains extensive discussion on relevant facts of Landau Level. We set $\hbar=c=1$

The Hamiltonian for an electron in a uniform magnetic field is
\begin{equation}
H=\frac{1}{2m_e}(\bm{p}+e\bm{A})^2
\end{equation}
where $m_e$ is the mass of electron. We take the magnetic field to point in the minus z direction, $\nabla \times \bm{A}=-B\hat{z}$, with $B>0$ and $e>0$.

The guiding center coordinates of the electrons are defined as
\begin{equation}
\bm{R}=\bm{r}+l^2 \bm{\pi}\times \hat{z}
\end{equation}
where $l^2=\frac{1}{eB}$ is the magnetic length, $\bm{r}$ is the position operator of electrons, $\bm{\pi}=\bm{p}+e\bm{A}$ is the kinetic momentum operator. It can be verified the following set of commutation relations hold among these operators ($i,j=x,y$)
\begin{equation}
[R_x,R_y]=-il^2\quad [R_i,\pi_j]=0\quad [r_i,\pi_j]=i\delta_{ij}
\end{equation}

The Hamiltonian can be written as
\begin{equation}
H=\omega_c(\pi^{\dagger}\pi+\frac{1}{2})\quad [\pi,\pi^{\dagger}]=1
\end{equation}
where we define $\omega_c=\frac{eB}{m_e}$,$\pi\equiv l\frac{\pi_x+i\pi_y}{\sqrt{2}}$.
It follows that the eigenvalues of $H$ are given by the Landau levels $E_n=\omega_c (n+\frac{1}{2})$

Since $H$ is only a function of $\bm{\pi}$, $\bm{R}$ is a constant of motion.
We now define the magnetic translation operators $t(\bm{d})$, which is the analog of ordinary translation operators in problems without magnetic field.
\begin{equation}
t(\bm{d})=\exp(i\bm{P}\cdot\bm{d})
\end{equation}
where $\bm{P}$ is defined via $\bm{R}$ as
\begin{equation}
\bm{P}\equiv \frac{1}{l^2} \hat{z}\times \bm{R}
\end{equation}

Using the Baker–Campbell–Hausdorff (BCH) formula, we can verify the following algebraic relation among the magnetic translation operators
\begin{equation}
\begin{split}
t(\bm{d_1})t(\bm{d_2})&=\exp(i(P_x d_{1x}+P_y d_{1y}))\exp(i(P_x d_{2x}+P_y d_{2y}))\\
&=\exp(i\bm{P}\cdot(\bm{d}_1+\bm{d}_2))\exp(\frac{i}{2l^2}\hat{z}\cdot(\bm{d_1}\times \bm{d_2}))\\
t(\bm{d_1})t(\bm{d_2})&=t(\bm{d_2})t(\bm{d_1})\exp(i\frac{i}{l^2}\hat{z}\cdot (\bm{d}_1\times\bm{d}_2))
\end{split}
\end{equation}
Thus, $t(\bm{d_1})$ commutes with $t(\bm{d_2})$ if the parallelogram spanned by $\bm{d_1}$ and $\bm{d_2}$ encloses an integer number of flux quanta, where each flux quanta occupies an area of $2\pi l^2$. 

This introduces the concept of the magnetic lattice on which  finite-size Landau Level problems are studied (e.g. in exact diagonalization studies of the Landau level). It describes a torus (or parallelogram with identified edges) that has an area of $N_1N_22 \pi l^2$.
It consists of lattice vectors $ n_1\bm{a}_1+n_2\bm{a_2}$, for $n_1\in [0,N_1-1]$, $n_2\in [0,N_2-1]$, where $|\bm{a_1}\times \bm{a_2}|=2\pi l^2$.
If we restrict $\bm{d}$ to being the lattice vectors, then $t(\bm{d})$ commute among themselves. 

Since $\bm{\pi}$ and $\bm{R}$ mutually commute, we may define the tensor product Hilbert space $\ket{n,\bm{k}}=\ket{n}\otimes\ket{\bm{k}}$.
Here, $\pi^\dagger \pi \ket{n}=n\ket{n}$ is the number basis of the Landau level index, and $\ket{\bm{k}}$ are eigenstates of the magnetic translation operators satisfying
% Therefore, we can introduce simultaneous eigenstates of the magnetic translation operators and Hamiltonian, i.e. magnetic Bloch states, labeled by $\ket{n,\bm{k}}$. We define them to obey the following relation 
\begin{equation}
\begin{split}
t(\bm{a_i})\ket{\bm{k}}&=\exp(i\phi_i)\exp(i\bm{k}\cdot\bm{a_i})\ket{\bm{k}}\\
% H\ket{n,\bm{k}}&=(n+\frac{1}{2})\omega_c\ket{n,\bm{k}}
\end{split}
\end{equation}
where $i=1,2$ and $\phi_1=\phi_2=\pi$\cite{PhysRevLett.127.246403}. 
The allowed \{$\bm{k}$\}  are determined by the following relation, 
\begin{equation}
\exp(i\bm{k}\cdot N_1\bm{a}_1)=1\quad \exp(i\bm{k}\cdot N_2\bm{a}_2)=1
\end{equation}
and they are defined modulo the magnetic RL vectors $\bm{g}_i$ ($\bm{g_i}\cdot \bm{a}_j=2\pi \delta_{ij}$).

\subsection{Form factors}
In the above, we restricted $\bm{k}$ of the state $\ket{\bm{k}}$ to be within the first magnetic BZ. Physically, the state $\ket{\bm{k}}$ must be equivalent to $\ket{\bm{k}+\bm{g_i}}$ up to a phase.  We now fix this phase. 
We define the operator $\lceil \bm{x} \rceil$ as sending $\bm{x}$ back to the first magnetic BZ, and define $\bm{g}_{\bm{x}} \equiv \bm{x}-\lceil \bm{x} \rceil$ to be the RL part of the vector $\bm{x}$ (distinct from the RL basis vectors $\bm{g}_i$).

We start with the state $\ket{\bm{0}}=\ket{k_x=0,k_y=0}$, and define all the other states as 
\begin{equation}
\ket{\bm{k}}\equiv\exp(i\bm{k}\cdot\bm{R})\ket{\bm{0}}
%= \exp(i\bm{k}\cdot \bm{R})\ket{\bm{0}}\equiv \ket{n}\otimes \ket{\bm{k}}
\end{equation}
where $\bm{k}$ is unrestricted. 
% The last step should be taken as definition of $\ket{n}$ and $\ket{\bm{k}}$.
This definition is consistent with what we discussed in the preliminary, since we can verify that $\ket{\bm{k}}$ defined in this way has the right eigenvalue $-\exp(i\bm{k}\cdot\bm{d})$ under magnetic translation operators $t(\bm{d})$.
% , and also has the right eigenvalue $(n+\frac{1}{2})\omega_c$ under the action of $H$. So, $\ket{n,\bm{k}}$ defined in this way is a magnetic Bloch state.

Under this definition of $\ket{\bm{k}}$, we can use the BCH formula to obtain
\begin{equation}
\exp(i\bm{g_i}\cdot \bm{R})\ket{\bm{k}}=\exp(\frac{il^2}{2}\hat{z}\cdot (\bm{g_i}\times \bm{k}))\ket{\bm{k}+\bm{g}_i}
\end{equation}
 At the same time, by using $\bm{g}_i= \epsilon_{ij}\hat{z}\times \bm{a}_j/l^2$, we have
\begin{equation}
\exp(i\bm{g_i}\cdot \bm{R})\ket{\bm{k}}=t(-l^2\bm{g}_{i}\times \hat{z})\ket{\bm{k}}=-e^{il^2\hat{z}\cdot (\bm{g_i}\times \bm{k})}\ket{n,\bm{k}}
\end{equation}
Thus, this fixes the phase between $\ket{\bm{k}+\bm{g}_i}$ and $\ket{\bm{k}}$.

\begin{equation}
\ket{n,\bm{k}+\bm{g}_i}=-e^{\frac{il^2}{2}\hat{z}\cdot (\bm{g_i}\times \bm{k})}\ket{n,\bm{k}}
\end{equation}

Let us now calculate the matrix element $\braket{\bm{k}_1|\exp(i\bm{q}\cdot \bm{R})}{\bm{k}_2}$, where $\bm{k_1}$ is in the BZ, but $\bm{k_2}$ is unrestricted. 
This quantity appears frequently in exact diagonalization studies of the Landau level.
\begin{equation}
\begin{split}
\braket{\bm{k}_1}{\exp(i\bm{q}\cdot \bm{R})|\bm{k}_2}
&=\braket{\bm{k}_1}{\bm{k}_2+\bm{q}}\exp(-\frac{il^2}{2}\hat{z}\cdot(\bm{q}\times\bm{k}_2))\\
&=\braket{\bm{k}_1}{\lceil \bm{k}_2+\bm{q}\rceil }\exp(-\frac{il^2}{2}\hat{z}\cdot(\bm{q}\times\bm{k}_2))\exp(\frac{il^2}{2}\hat{z}\cdot(\bm{g}_{\bm{k_2}+\bm{q}}\times \lceil \bm{k}_2+\bm{q}\rceil))\eta(\bm{g}_{\bm{k_2}+\bm{q}})\\
&=\delta_{\bm{k_1},\lceil \bm{k}_2+\bm{q}\rceil}\exp(-\frac{il^2}{2}\hat{z}\cdot(\bm{q}\times\bm{k}_2))\exp(\frac{il^2}{2}\hat{z}\cdot(\bm{g}_{\bm{k_2}+\bm{q}}\times \lceil \bm{k}_2+\bm{q}\rceil))\eta(\bm{g}_{\bm{k_2}+\bm{q}})
\end{split}
\end{equation}
where $\eta(\bm{x})$ is -1 if $\frac{\bm{x}}{2}$ is not on the magnetic RL, and 1  if it on the magnetic RL.

For the form factors, we want to calculate 
\begin{equation}
% \braket{u_{m,\bm{k}}}{u_{n,\bm{k}-\bm{q}}}=
F_{mn}(\bm{k},\bm{k}-\bm{q})=
% \braket{u_{m,\bm{k}}}{u_{n,\bm{k}-\bm{q}}}=
\braket{m,\bm{k}}{\exp(i\bm{q}\cdot \bm{r})|n,\bm{k}-\bm{q}}
\end{equation}
 where $\bm{q}$ is unrestricted, and $m,n$ are Landau level indices.
We can decompose $\bm{r}=\bm{R}+l^2{\hat{z}\times \bm{\pi}}$, then
\begin{equation}
F_{mn}(\bm{k},\bm{k}-\bm{q})=\braket{m}{\exp(il^2\bm{q}\cdot (\hat{z}\times\bm{\pi}))|n} \braket{\bm{k}}{\exp(i\bm{q}\cdot \bm{R})|\bm{k}-\bm{q}}
\end{equation}
 We have already calculated the second term of the right hand side. 
 For the first term, since
\begin{equation}
\ket{n}=\frac{(\pi^{\dagger})^n}{\sqrt{n!}}\ket{0},\quad il^2 \bm{q}\cdot (\hat{z}\times \bm{\pi})=\frac{l}{\sqrt{2}}(\pi^{\dagger}q-\pi q^*)
\end{equation}
where $q\equiv q_x+iq_y$,
% The real space representation of $\ket{0,\bm{k}}$, i.e. the lowest Landau level wave function on a torus, is given by the modified Weierstrass sigma-function\cite{Haldane_2018_modifed,Wang2021exact}. 
% Evaluating $\braket{m}{\exp(il^2\bm{q}\cdot (\hat{z}\times\bm{\pi}))|n}$ just involves standard harmonic ladder operator algebra,
we have that
\begin{equation}
% \begin{split}
\braket{m}{\exp(il^2\bm{q}\cdot (\hat{z}\times\bm{\pi}))|n}=
\begin{cases}
(l(q_x+iq_y)/\sqrt{2})^{m-n}\sqrt{\frac{n!}{m!}}L_{n}^{m-n}(l^2\frac{|q|^2}{2})\exp(-l^2|q|^2/4)&\quad  (n\leq m)\\
  (-l(q_x-iq_y)/\sqrt{2})^{n-m}\sqrt{\frac{m!}{n!}}L_{m}^{n-m}(l^2\frac{|q|^2}{2})\exp(-l^2|q|^2/4) &\quad (n\geq m)
% \end{split}
\end{cases}
\end{equation}
Where $L^{m}_{n}$(x) is associated Laguerre polynomial. Taking $n=m=0$, we obtain the form factor for the lowest Landau level,
\begin{equation}
F_{00}(\bm{k},\bm{k}-\bm{q})=\exp(-\frac{l^2 |q|^2}{4})\exp(-i\frac{l^2}{2}\hat{z}\cdot (\bm{q}\times \bm{k}))
\end{equation}
This matches exactly with the form factors given in the main text after the identification of $l^2\to \mathcal{B}$

\begin{figure}[t]
\begin{center}
\includegraphics[width=0.9\textwidth]{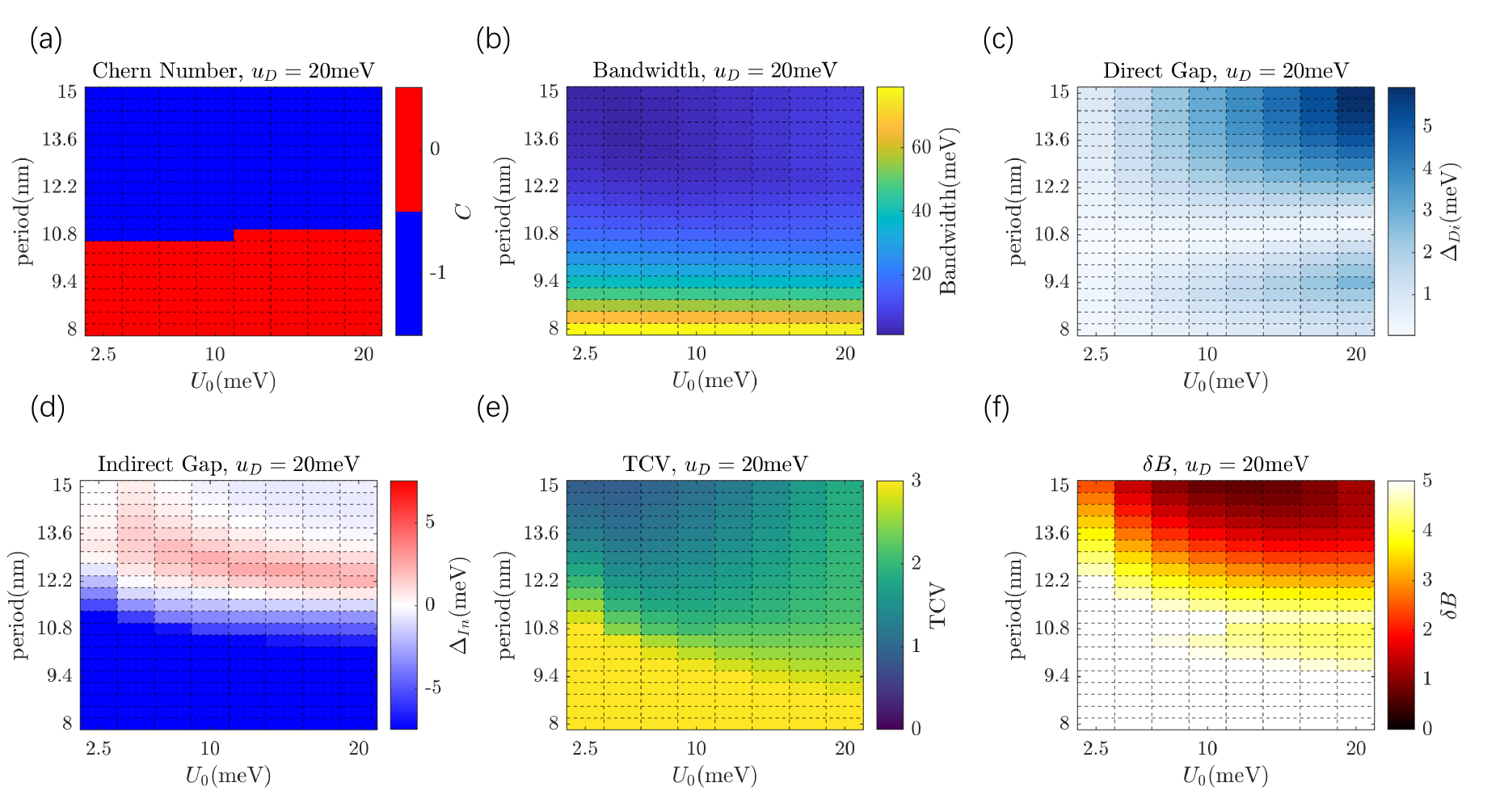}
\end{center}
\caption{Properties of the first conduction band when R5G is subject to periodic scalar potential $U(\bm{r})$ at displacement field $u_D=20$meV. The (a) Chern number, (b) bandwidth, (c) direct gap, (d) indirect gap, (e) trace condition violation (TCV), and (f)Berry curvature variance $\delta B$ is shown.
}
\label{SPfig_uD20}
\end{figure}

\begin{figure}[t]
\begin{center}
\includegraphics[width=0.9\textwidth]{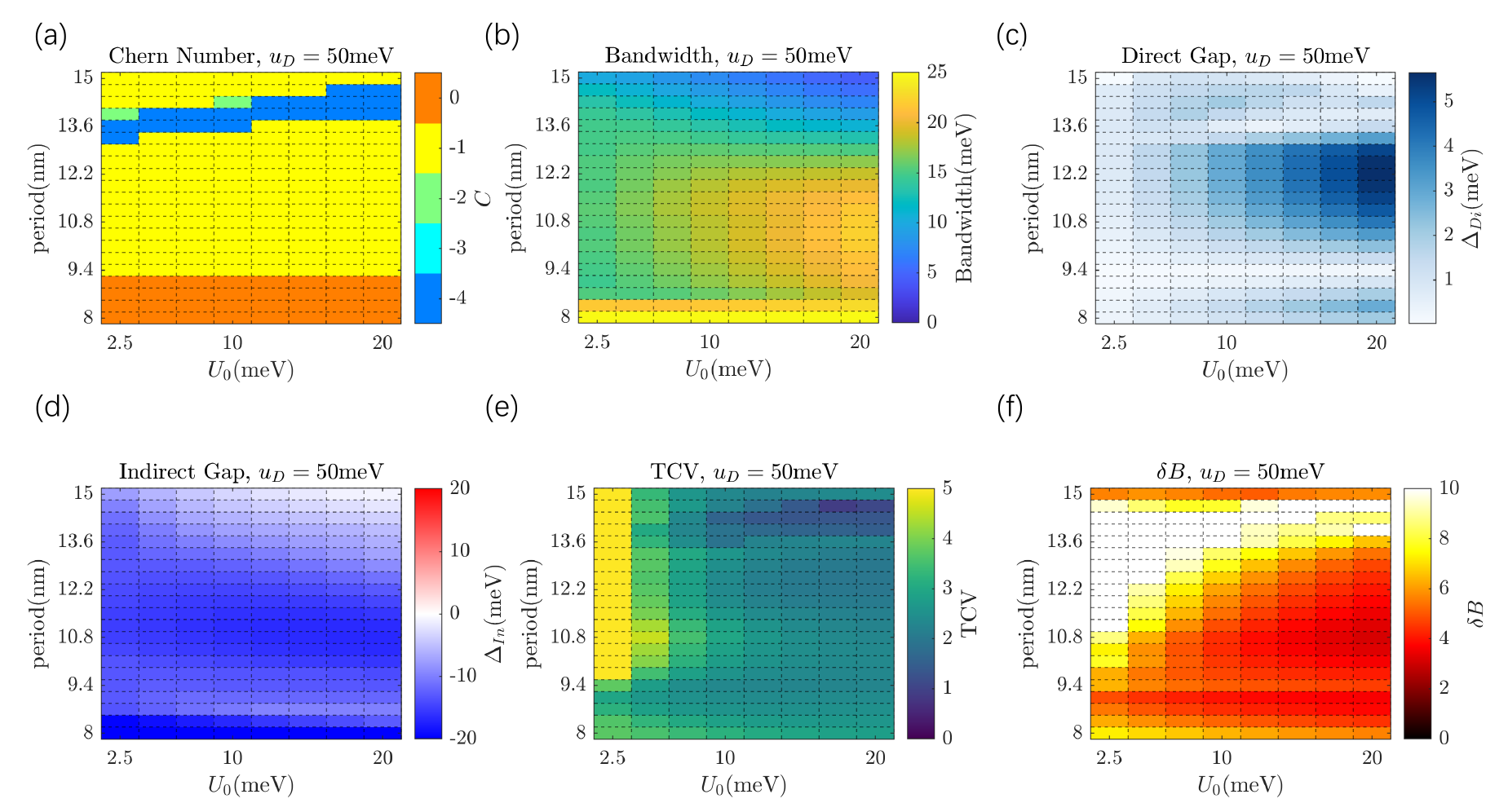}
\end{center}
\caption{
Same as Fig~\ref{SPfig_uD20} but for $u_D=50$meV.
% Properties of the first conduction band when R5G is subject to $V_{\text{po}}(\vec{r})$ at displacement field  $u_D=50$meV/layer. (a) Chern Number (b) Bandwidth (c)Direct gap (d) Indirect gap (e) Trace-condition violation TCV (f)Berry curvature variance $\delta B$
}
\label{SPfig_uD50}
\end{figure}

\section{Rhombohedral Pentalayer Graphene details}

The methodology of doing Hartree-Fock calculation in pentalayer graphene (R5G)+hBN is described abundantly in the literature \cite{dong2023anomalous,kwan2023moir,dong2024stability,zhou2023fractional,guo2023theory}. We include necessary information for reproducing our calculation presented in the main text. We follow closely the convention of Ref.\cite{dong2023anomalous}.~\footnote{The authors acknowledge useful private conversion with Y. Kwan and J. K. Dong on the techniques.}
The geometry of R5G on hBN and the associated single-particle moir\'e Hamiltonian are detailed in Ref.\cite{dong2023anomalous} and reference therein \cite{PhysRevB.82.035409,PhysRevB.88.075408,PhysRevLett.132.116504,PhysRevB.90.155406}. We copy some necessary information here for the purpose of being self-contained.

\subsection{Single particle Hamiltonian}
% The Hartree-Fock calculation presented in the main text ignores hBN. 
% We discuss the set up in that scenario first
We first introduce the single-particle Hamiltonian of R5G.
The direct atomic lattice of R5G is spanned by
\begin{equation}
\vec{R}_1=a_{\text{Gr}}(1,0)\quad \vec{R}_2=a_{\text{Gr}}(\frac{1}{2},\frac{\sqrt{3}}{2})
\end{equation}
where $a_{\text{Gr}}=0.246 \text{nm}$ is the lattice constant of graphene. The reciprocal lattice of R5G is spanned by
\begin{equation}
\vec{G}_1=\frac{4\pi}{\sqrt{3}a_{\text{Gr}}}(\frac{\sqrt{3}}{2},-\frac{1}{2})\quad \vec{G}_2=\frac{4\pi}{\sqrt{3}a_{\text{Gr}}}(\frac{1}{2},\frac{\sqrt{3}}{2})
\end{equation}

The Hamiltonian of $N_L$-layer pristine rhombohedral graphene is given by~\cite{dong2023anomalous}
\begin{equation}
H_{RG}=\sum_{k\in BZ}c^{\dagger}_{k,\sigma,l} [h_{RG}(\vec{k})]_{\sigma,l;\sigma',l'}c_{k,\sigma',l'}
\end{equation}
where $\sigma$ labels sublattice, $l=1..N_L$ labels layer, and $c^{\dagger} (c)$ are fermion creation (annihilation) operators.
The intralayer ($l=l^\prime=L$) Hamiltonian is given by
\begin{equation}
[h_{RG}(\vec{k})]_{\sigma,L;\sigma',L}=
\begin{pmatrix}
u_L & -t_0f(\vec{k})\\
-t_0 f^*(\vec{k}) & u_L
\end{pmatrix}_{\sigma\sigma'}
\end{equation}
The non-zero interlayer tunneling Hamiltonians are given by
\begin{equation}
[h_{RG}(\vec{k})]_{\sigma,L;\sigma',L+1}=
\begin{pmatrix}
t_4 f(\vec{k}) & t_3 f^*(\vec{k})\\
t_1 & t_4 f(\vec{k})
\end{pmatrix}_{\sigma\sigma'}
\end{equation}
\begin{equation}
[h_{RG}(\vec{k})]_{\sigma,L;\sigma',L+2}=
\begin{pmatrix}
0 & t_2\\
0 & 0
\end{pmatrix}_{\sigma\sigma'}
\end{equation}
where
\begin{equation}
f(\vec{k})=\sum_{i=1}^3 \exp(i\vec{k}\cdot \vec{\delta_i})
; \;\;\;\;\;
\vec{\delta}_i=R_{\frac{i 2\pi}{3}}(0,\frac{a_{Gr}}{\sqrt{3}})^T
\end{equation}
In the above, $u_L=(L-\frac{(1+N_L)u_D}{2})$ is the onsite potential on the $L$th layer introduced by the displacement field ($L=1..N_L$), and $(t_0,t_1,t_2,t_3,t_4)=(3100,380,-21,290,141)$meV are hopping parameters for graphene~\cite{PhysRevB.82.035409}.  
% $\vec{\delta}_i$ are vectors pointing from the carbon atom toward its three nearest neighbours.

\subsection{Periodic scalar potential}
In the main text, we presented the single-particle phase diagram when R5G is subject to a scalar moir\'e potential that acts equally on each layer
\begin{equation}
U(\vec{r})=-2U_0[ \cos(\vec{g}_1\cdot \vec{r}+\phi)+\cos(\vec{g}_2\cdot \vec{r}+\phi)+\cos((\vec{g}_2+\vec{g}_1)\cdot \vec{r}+\phi)]
\end{equation}
where
\begin{equation}
\vec{g}_1=\frac{4\pi}{\sqrt{3}a}(\frac{\sqrt{3}}{2},-\frac{1}{2})\quad \vec{g}_2=\frac{4\pi}{\sqrt{3}a}(\frac{1}{2},\frac{\sqrt{3}}{2})
\end{equation}
are the reciprocal lattice vectors of the periodic potential.
Under this scalar potential, the bands of R5G will fold and split into minibands.
% where $\vec{g}_{1}$ ($\vec{g}_{2}$) points in the direction of $\vec{G}_1$ ($\vec{G}_2$), but with magnitude $|\vec{g}_{i}|=\frac{4\pi}{\sqrt{3}a}$, with $a=11$nm and $\phi=\pi$ as discussed in the main text. 
In the main text, we choose $a=11$nm and $\phi=\pi$ to maximize the $C=-1$ gap.

Here, we show additional details of these minibands.
The properties of the first conduction miniband as a function of potential strength $U_0$ and period $a$ is shown in Fig.~\ref{SPfig_uD20} for $u_D=20$meV and in Fig.~\ref{SPfig_uD50} for $u_D=50$meV. 
As can be observed in the first panel of these two figures, the $C=-1$ phase is quite dominant in the relevant parameter region, consistent with the results in the uniform Berry curvature limit.  %ideal parent band  under a scalar potential.

\subsection{Hartree-Fock in the band basis}\label{sec:pentalayerhf}

For the Hartree-Fock calculation, we  allow translation symmetry to be broken with the same lattice vectors as in the scalar potential case above. Note that there are no single-particle terms that explicitly break the translational symmetry. 
By choosing this set of reciprocal lattice vectors, we are allowing the interactions to spontaneously break  translation symmetry according to the chosen pattern.
In this calculation, we assume valley polarization in the $\vec{K}=(\frac{4\pi}{3a_{Gr}},0)$ valley, and only consider one spin species in this valley. 
For notational simplicity, we omit the arrow symbol of the vector in this subsection.

The main text calculation is carried out by projecting the Hamiltonian to the first $N_b$ conduction minibands of the single-particle R5G Hamiltonian folded according to the chosen $g_i$. We set $N_b=7$ throughout our calculation. We denote the miniband creation operators by $c_{k,\alpha}^{\dagger}$, where $\alpha=1...N_b$. It creates an electron in the Bloch eigenstate of the single-particle R5G moir\'e Hamiltonian, denoted below as $\ket{k\alpha}$ with energy $\epsilon_{k,\alpha}$. The projected many-body Hamiltonian can be written as
\begin{equation}
H=\sum_{\alpha,k\in BZ}\epsilon_{k,\alpha}c_{k,\alpha}^{\dagger}c_{k,\alpha}+\frac{1}{2}\sum_{k_1,k_2,k_3,k_4\in \mathrm{BZ}}^{\alpha,\beta,\gamma,\delta}V_{k_1\alpha,k_2\beta;k_2\gamma,k_1\delta} c^{\dagger}_{k_1,\alpha}c^{\dagger}_{k_2,\beta}c_{k_4\delta}c_{k_3\gamma}
\end{equation}
The Coulomb matrix element is 
\begin{equation}
V_{k_1\alpha,k_2\beta;k_2\gamma,k_1\delta}=\bra{k_1\alpha,k_2\beta}\hat{V}\ket{k_3\gamma,k_4\delta}=\frac{1}{A}\sum_{q}V(q)\braket{k_1\alpha}{\exp(-i\vec{q}\cdot \vec{r})|k_3\gamma}\braket{k_2\beta}{\exp(i\vec{q}\cdot \vec{r})|k_4\delta}
\end{equation}
where the summation of $q$ is over all possible momentum transfer, $V(q)$ is the Fourier transform of the Coulomb potential, and $A$ is the sample area. Here we take dual-gate screened Coulomb potential with gate-to-sample distance $d=25$nm. 

\begin{equation}
V(q)=\frac{e^2 \text{tanh}(|q|d)}{2\epsilon_r\epsilon_0 |q|}
\end{equation}

On a finite-size system, $k_i$ is summed over discrete momentum vectors (the ``mesh'') in the BZ, and $q$ is over all the discrete momentum vectors in the mesh plus RL vectors up to some large cutoff, taken to be $|q|_{\text{max}}=6|g_i|$. 
The mesh is taken to be  
\begin{equation}
k_i=\frac{n_1}{N_q}g_1+\frac{n_2}{N_q}g_2
\end{equation}
where $n_1,n_2=0..N_q-1$. We take $N_q=15$ in our calculation.

Before proceeding, let us define the form factors $\Lambda$
\begin{equation}
\Lambda_{q}(k_1)_{\alpha,\beta}=\braket{u_{k_1\alpha}}{u_{k_1+q\beta}}=\braket{k_1,\alpha}{e^{-iqr}|k_2,\beta}\delta_{k_1,[k_2-q]}
\end{equation}
In this definition, we restrict $k_1,k_2$ to be within the mBZ, but $q$ is unrestricted. $\ket{u_{k,\alpha}}=e^{-ik\cdot r}\ket{k,\alpha}$ is the periodic part of the Bloch wave function $\ket{k,\alpha}$. We have chosen a periodic gauge such that $\ket{k,\alpha}=\ket{k+g_i,\alpha}$, i.e. the Bloch wave function is periodic under shift by RL. In practice, we only need to compute $\Lambda$  one time before the start of  the Hartree-Fock iteration by using the single-particle R5G(+hBN) moir\'e Hamiltonian wave functions. 

The Hartree-Fock approximation consists of substituting the four-fermion operator by its contractions in the projected many-body Hamiltonian $H$
\begin{equation}
\begin{split}
c^{\dagger}_{k_1,\alpha}c^{\dagger}_{k_2,\beta}c_{k_4\delta}c_{k_3\gamma}\to&
(\langle c^{\dagger}_{k_1,\alpha} c_{k_3,\gamma}\rangle c^{\dagger}_{k_2,\beta} c_{k_4,\delta}+ c^{\dagger}_{k_1,\alpha} c_{k_3,\gamma} \langle c^{\dagger}_{k_2,\beta} c_{k_4,\delta}\rangle)\delta_{k_2,k_4}\delta_{k_3,k_1}\\
&(-\langle c^{\dagger}_{k_1,\alpha} c_{k_4,\delta}\rangle c^{\dagger}_{k_2,\beta} c_{k_3,\gamma}-c^{\dagger}_{k_1,\alpha} c_{k_4,\delta} \langle c^{\dagger}_{k_2,\beta} c_{k_3,\gamma}\rangle)\delta_{k_2,k_3}\delta_{k_4,k_1}
\end{split}
\end{equation}
After this substitution,  the interacting part of the Hamiltonian can be divided into Hartree part and Fock part
\begin{equation}
\begin{split}
H_{H,\beta\delta}(k_2)&=\frac{1}{A}\sum_{\substack{k_1\in \mathrm{BZ}\\ q \in \mathrm{RL}\\\alpha,\gamma}} V(q)P_{\gamma,\alpha}(k_1)\Lambda_{g}(k_2)_{\beta\delta}(\Lambda_g(k_1)_{\gamma\alpha})^*\\
&=\frac{1}{A}\sum_{q\in \mathrm{RL}}V(q)\Lambda_{q}(k_2)_{\beta\delta}[\sum_{k_1\in \mathrm{BZ}} \text{Tr}\{P(k_1)\Lambda_{q}^{\dagger}(k_1)\} ]
\end{split}
\end{equation}
\begin{equation}
\begin{split}
H_{F,\beta\gamma}(k_2)&=\frac{-1}{A}\sum_{\substack{k_1 \in \mathrm{BZ}\\q\in \lceil k_1-k_2 \rceil+\mathrm{RL}\\ \alpha,\delta}}V(q)\Lambda_{q}(k_2)_{\beta\delta}P_{\delta\alpha}(k_1)(\Lambda_q(k_2)_{\gamma\alpha})^*\\
&=\frac{-1}{A}\sum_{\substack{k_1 \in \mathrm{BZ}\\q\in \lceil k_1-k_2 \rceil+\mathrm{RL}}}V(q)[\Lambda_q(k_2)P(k_1)\Lambda^{\dagger}_{q}(k_2)]_{\beta\gamma}
\end{split}
\end{equation}
where we have defined the projector 
\begin{equation}
{P}_{\alpha\beta}(k)=\langle c^{\dagger}_{k\beta}c_{k\alpha} \rangle
%=\sum_{n\in occ} z^{n}_{k\alpha}z^{n*}_{k\beta}
\end{equation}
and $\langle\dots \rangle$ is with respect to the Slater determinant ground state. 

% The Hartree-Fock orbitals are expressed in terms of the band operators
% \begin{equation}
% d^{\dagger}_{kn}=\sum_{\alpha}z_{k\alpha}^{n}c^{\dagger}_{k,\alpha}
% \end{equation}
% where $n=1..N_b$ labels the (Hartree-Fock) band.
% The Slater determinant ground state is 
% \begin{equation}
% \ket{\mathrm{Slater}}=\prod_{n,k \in \mathrm{filled}}d^{\dagger}_{k,n}\ket{\Omega}
% \end{equation}

Notice that our calculation is carried out by projecting to the conduction band, and there is no subtraction associated with the projector ${P}$. This is justified when the band gap ($u_D$) is large, and in the case without hBN where the background charge density is uniform, shown in the main text.
Subtleties associated with the subtraction in ${P}$ are discussed extensively in Ref.\cite{kwan2023moir}.

The full Hartree-Fock Hamiltonian is
\begin{equation}
H_{HF}=\sum_{\alpha,k\in \mathrm{BZ}}\epsilon_{\alpha,k}c_{k,\alpha}^{\dagger}c_{k,\alpha}+\sum_{k,\alpha,\beta} (H_{H,\alpha\beta}(k)+H_{F,\alpha\beta}(k))c^{\dagger}_{k\alpha}c_{k\beta}
\end{equation}
The self-consistent HF equations are solved iteratively until self-consistency is achieved, determined by convergence of the projector $P$.
% The Hartree-Fock orbitals $d^{\dagger}_{kn}$ are obtained by diagonalizing $H_{HF}$, which give rise to the new projector $\mathcal{P}$ and a new $H_{HF}$.  The procedure is repeated until self-consistency is achieved, determined by convergence of $\mathcal{P}$ and the energy.
% We use a random initial ${P}$ to locate the lowest-energy state.

% The Hartree-Fock orbitals $d^{\dagger}_{kn}$ are obtained by diagonalizing $H_{HF}$. The coefficients $z's$ are the eigenvectors of it. $H_{HF}$ depends on the projector $\mathcal{P}$, and $\mathcal{P}$ is constructed from the eigenvectors of $H_{HF}$. The process is repeated until convergence of $\mathcal{P}$ (and energy). We use a random initial $\mathcal{P}$ to locate the lowest energy states.

The energy of the final state is 
\begin{equation}
E=\sum_{k}\text{Tr}[(H_{\text{single}}(k)+\frac{H_{H}(k)+H_{F}(k)}{2})P(k)]
\end{equation}
where $H_{\text{single}}(k)_{\alpha,\beta}=\epsilon_{k,\alpha}\delta_{\alpha\beta}$

\subsection{The effect of hBN}

For completeness, we show results with the effect of hBN relevant to the experiment~\cite{lu2024fractional}.
The direct lattice of the hBN is spanned by
\begin{equation}
\vec{R}^{\text{hBN}}_i=\frac{a_{\text{hBN}}}{a_{\text{Gr}}}R_{\theta}\vec{R}_i
\end{equation}
where $a_{\text{hBN}}=0.2504$nm is the lattice constant of hBN, and $R_{\theta}$ is the counter-clockwise rotation by $\theta$. 
The moir\'e RL basis vectors of the R5G+hBN system are 
\begin{equation}
\vec{g}_{i}=\left(I-\frac{a_{\text{Gr}}}{a_{\text{hBN}}}R_{\theta}\right)\vec{G}_i
\end{equation}

The effect of the hBN is modeled by adding 
\begin{equation}
V_{hBN}=V_0 \sigma_0+V_1 \vec{N}(\vec{r})\cdot \vec{\sigma}
\end{equation}
where
 to the bottom layer of the R5G Hamiltonian.  Here, $\sigma$ are Pauli matrices acting on the sublattices of the bottom layer, and
\begin{equation}
\vec{N}(\vec{r})\cdot \vec{\sigma}=
\exp(-i\psi)[\exp(i\vec{g_1}\cdot \vec{r})\begin{pmatrix}
1&1\\
\omega&\omega
\end{pmatrix}+\exp(i\vec{g_2}\cdot \vec{r})\begin{pmatrix}
1&\omega^*\\
\omega^*&\omega
\end{pmatrix}+\exp(-i(\vec{g_1}+\vec{g_2})\cdot \vec{r})\begin{pmatrix}
1&\omega\\
1&\omega
\end{pmatrix}
+h.c.]
\end{equation}
where $\omega=\exp(i\frac{2\pi}{3})$, and $(V_0,V_1,\psi)$ are parameters determining the coupling strength of hBN with R5G.  From first-principles,
$(V_0,V_1,\psi)=(28.9\text{meV},21\text{meV},-0.29)$\cite{dong2023anomalous,PhysRevB.90.155406}.

% In the recent experiment reporting FCI in R5G+hBN system, the rotation angle between the R5G and hBN substrate is $\theta\approx 0.77^{\circ}$, which corresponds to a moir\'e period of $11$nm in real space \cite{lu2024fractional}. This is the reason why we set $|\vec{g}_i|$ to be the corresponding magnitude even for calculation without hBN. 

\begin{figure}[t]
\begin{center}
\includegraphics[width=0.6\textwidth]{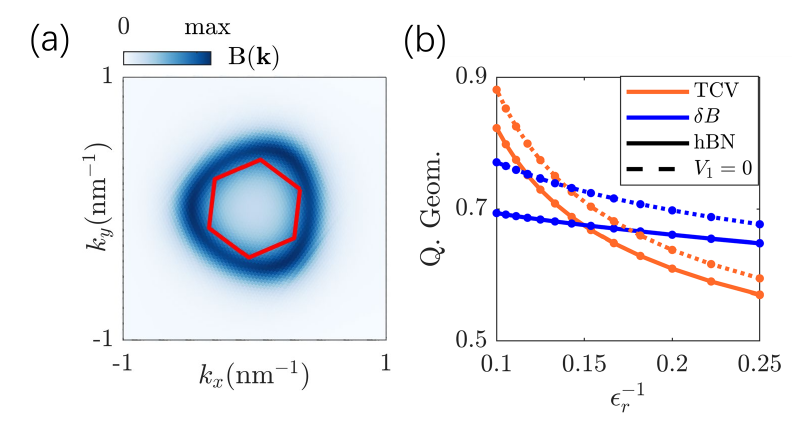}
\end{center}
\caption{ (a)Berry curvature of the rhombohedral pentalayer graphene near $K$ valley. The red line delineates the BZ corresponding to $\theta=0.77^{\circ}$, as used in (b). (b)Quantum geometrical quantities, trace condition violation TCV and Berry curvature standard deviation $\delta B$, of the first conduction of the pentalayer graphene under band projected Hartree-Fock calculation when the twist angle between hBN and graphene is $\theta=0.77^{\circ}$. The solid lines are for $(V_0,V_1,\psi)=(28.9\text{meV},21\text{meV},-0.29)$ while the dashed lines are for artificially turning off the moir\'e part of the hBN potential  $(V_0,V_1,\psi)=(28.9\text{meV},0\text{meV},-0.29)$
}
\label{SPfig_R5G}
\end{figure}

 We show the HF results when hBN is properly taken into account with the rotated BZ shown in Fig.\ref{SPfig_R5G} at $\theta=0.77^{\circ}$~\cite{lu2024fractional}. 
 As can be observed with a comparison with the main text, the proper inclusion of hBN does not qualitatively change the results.
 % stronger interaction  leads to improvement in the quantum geometry of AHC. This is expected, since with positive large $u_D$, the electrons in the conduction bands are localized on the layer far away from hBN, making its effects negligible.

\section{Connecting the spinor structure of rhombohedral multilayer graphene and the infinite Chern band}
The spinor for the infinite Chern band defined in the main text is
\begin{equation}
\ket{s_{\bm{k}}^{\mathcal{B}}}
=e^{\sqrt{\mathcal{B}}(k^*b^\dagger-kb)}\ket{0}
= e^{-\frac{\mathcal{B}}{4}|\bm{k}|^2} \sum_{n=0}^{\infty} \frac{(\sqrt{\mathcal{B}} k^*)^n}{\sqrt{n!}}\ket{n}
\end{equation}
We now show that a similar spinor structure also arises in rhombohedral multilayer graphene, and discuss a certain limit in which it is realized exactly.
% We will now discuss a limit in which this spinor also arises in the rhombohedral multilayer graphene system. 

Consider the rhombohedral multilayer graphene Hamiltonian with only interlayer-hopping $t_1$. 
This can be written as 
\begin{equation}
H_{\text{RMG}}(\bm{k})=\sum_{n=0}^{N_L-2}(-t_0 f^*(\bm{k})) B^{\dagger}_n A_n+t_1A_n^{\dagger}B_{n+1}+h.c.
\end{equation}
where $f(\bm{k})=\sum_i\exp(i\bm{k}\cdot \bm{\delta_i})$, $n$ labels the layer index (counting from the top), $N_L$ is the total number of layers, and $A_n^\dagger$ ($B_n^\dagger$) are creation operators for the state on sublattice $A$ ($B$) and layer $n$ (all at momentum $\bm{k}$).
% , i.e. the layer furthest from hBN will have the n=0. This is different from the convention we used in the Hartree-Fock calculation, but it proves convenient here. A/B labels sublattice. . 

When expanded near the $K$ point, $\bm{k}=(\frac{4\pi}{3a_{Gr}}-q_x,-q_y)$, we have
\begin{equation}
H_{\text{RMG}}(\bm{q})=\sum^{N_L}_{n=0}(-v_F)(q_x+iq_y)B^{\dagger }_nA_n+\sum^{N_L-1}_{n=0}t_1 A^{\dagger}_n B_{n+1}+h.c.
\end{equation}
where $v_F = \frac{\sqrt{3}}{2}a_{Gr} t_0$.
$H_{\mathrm{RMG}}$ is equivalent to the Su–Schrieffer–Heeger chain with the intra-cell hopping $-v_F(q_x+iq_y)$ and inter-cell hopping $t_1$. 
If we focus on the small $|\bm{q}|$ momentum region and in the limit of large layer number $N_L\rightarrow\infty$, the exact (unnormalized) zero-energy edge mode of $H_{\text{RMG}}(\bm{q})$ can be written 
\begin{equation}
\ket{E_{\bm{q}}}=\sum_{n=0}^{N_L-1}(q_x-iq_y)^{n}\left(\frac{v_F}{t_1}\right)^n B^{\dagger}_{n}\ket{\Omega}
\end{equation}
where $\ket{\Omega}$ is the vacuum.  
The edge mode lives entirely on the B sublattice and decays exponentially in the bulk.
This edge mode spinor, as well as its finite-layer version, was used in Ref.\cite{soejima2024anomalous}.

Notice that since $\ket{E_{\bm{q}}}$ is a holomorphic function of $q_x-i q_y$, up to a normalization constant, it must also describe a band satisfying the trace condition $\Tr g^{\mathrm{FS}}_E(\bm{q})- |B_E(\bm{q})| = 0$~\cite{wang2021exact}.
However, the Berry curvature distribution
\begin{equation}
B_E(\bm{q})=\frac{2(v_F/t_1)^2}{(|v_F \bm{q}/t_1|^2-1)^2} %add correct factors
\end{equation}
which is non-uniform (and diverges at $|\bm{q}|=t_1/v_F$, corresponding to the topological phase transition of the SSH chain beyond which the edge mode is no longer well defined).
However, notice the striking similarity between $\ket{E_{\bm{q}}}$ and the spinor $\ket{s_{\bm{k}}^{\mathcal{B}}}$ of the uniform parent band above.  

Consider a modified version of $H_{\text{RMG}}$ in which the interlayer tunneling is layer dependent by a factor of $\sqrt{n+1}$,
\begin{equation}
\tilde{H}_{\text{RMG}}(\bm{q})=\sum_{n=0}^{N_L-1}(-v_F(q_x+iq_y))B^{\dagger }_nA_n+\sum_{n=0}^{N_L-2}t_1\sqrt{n+1} A^{\dagger}_n B_{n+1}+h.c.
\end{equation}
This Hamiltonian has the exact zero-energy edge mode for large $N_L$,
\begin{equation}
\ket{\tilde{E}_{\bm{q}}}=\sum_{n=0}^{N_L-1}\frac{(\frac{q_x-iq_y}{\sqrt{2}})^{n}}{\sqrt{n!}}\left(\frac{\sqrt{2}v_F}{t_1}\right)^n B^{\dagger}_{n}\ket{\Omega}
\end{equation}
If we now identify $\frac{\sqrt{2}v_F}{t_1}\to \sqrt{\mathcal{B}}$, $\bm{q}\rightarrow\bm{k}$, and $B^{\dagger}_{l}\ket{\Omega}\to \ket{n}$, we recover exactly the spinor $\ket{s_{\bm{k}}^{\mathcal{B}}}$.
Thus, this modified RMG Hamiltonian gives rise to exactly the ideal parent band in the limit of large $N_L$.  
% For small $|\bm{k}|\$

 \begin{comment}
\begin{equation}
\tilde{H}_{\text{RMG}}(\bm{k})=\sum_{n=0}^{N_L}(-\frac{\sqrt{3}a_{Gr} t_0}{2})(q_x-iq_y)A^{\dagger }_nB_n+\sum_{n=0}^{N_L-1}t_1\sqrt{\frac{N_L-n}{n+1}} B^{\dagger}_n A_{n+1}+h.c.
\end{equation}
Where $N_L+1$ is the total number of layers. Under such a Hamiltonian, we can construct a the following (unnormalized) zero-energy edge-states of it (At finite $N_L$, this is not an exact eigenstate. But its overlap with the exact edge eigenstate becomes exponentially better at large $N_L$)
\begin{equation}
\ket{\psi}=\sum_{n=0}^{N_L}\sqrt{\frac{(N_L)!}{n!(N_L-n)!}}(\frac{2t_1}{(\sqrt{3}a_{Gr}t_0)(q_x-iq_y)})^n B^{\dagger}_n\ket{\Omega}
\end{equation}

If we now identify $\frac{2t_1}{\sqrt{3}a_{Gr}t_0}\to M$, $N_L\to 2S$, $n \to S-\sigma$, $B^{\dagger}_n\ket{\Omega}\to \ket{\sigma}$  we recover the ideal-model spinor (up to an inessential overall phase). Notice that $\ket{\psi}$ is localized on the the large $n$ layers for small $|\bm{q}|$, and this is consistent with the fact that under positive displacement field, the conduction band electron that hosts FCI is localized on the large $n$ layers \cite{dong2023anomalous}.
\end{comment}

\section{Energy of the Gaussian ansatz}
In this section, we calculate the kinetic and Fock energy of the Gaussian ansatz wavefunction in the limit of large $\xi$ at $\mathcal{B}=2\pi N/\Omega_{\mathrm{BZ}}$.
Because the $\hat{Q}$ mapping is exact for the Hamiltonian $\hat{\mathcal{H}}_0+\hat{\mathcal{H}}_F$, rather than calculating the energy for the non-trivial parent band, we can take advantage of the mapping and calculate the corresponding kinetic and Fock energies for the trivial parent band.

The ansatz wavefunction is
\begin{equation}
\psi_{\bm{k}}(\bm{r}) = \frac{1}{\sqrt{N_{\mathrm{uc}}}}\sum_{\bm{R}} e^{i\bm{k}\cdot\bm{R}} \phi_{\bm{R}}(\bm{r})
\end{equation}
where $N_{\mathrm{uc}}$ is the total number of unit cells, and
\begin{equation}
\phi_{\bm{R}}(\bm{r}) = \sqrt{\frac{2\xi^2}{\pi}}e^{-\xi^2|\bm{r}-\bm{R}|^2}.
\end{equation}
This state describes a lattice of Gaussian localized charges at lattice positions $\bm{R}$.
We are interested in the large $\xi$ limit, in which the overlap between Gaussians becomes negligible.
% The neglected terms, which arise from intersite Gaussian overlaps, are strongly suppressed by the Gaussian factor.
In this limit, the non-orthogonality of $\phi_{\bm{R}}(\bm{r})$ for different $\bm{R}$ becomes unimportant, and
$\psi_{\bm{k}}$ is a properly normalized wavefunction (away from this limit, there will be a $\bm{k}$-dependent normalization factor due to non-orthogonality of the individual $\phi_{\bm{R}}(\bm{r})$).
We consider the Slater determinant obtained by filling all $\psi_{\bm{k}}(\bm{r})$ states.

At large $\xi$, where the overlap of nearby Gaussians is zero, the kinetic energy per electron can be obtained from a single Gaussian at (say) $\bm{R}=\bm{0}$,
\begin{equation}
\begin{split}
E_K &= \int d\bm{r} \phi_{\bm{0}}^*(\bm{r}) \frac{-\nabla^2}{2m} \phi_{\bm{0}}(\bm{r}) = \frac{\xi^2}{m}
\end{split}
\end{equation}

Next, we consider the Fock energy.
Although the overlap between neighboring Gaussians is effectively zero, this does not mean that the Fock, or ``exchange'', energy is zero.
This is because in the momentum space Hartree-Fock decomposition of the energy, the Hartree energy contains an unphysical self-interaction term that must be canceled out by the Fock term.
To see this,
let $\rho(\bm{r}) = \sum_{\bm{R}}|\phi_{\bm{R}}(\bm{r})|^2$ be the total background charge density.
Then, the Hartree energy per electron,
\begin{equation}
E_H = \frac{1}{N_{\mathrm{uc}}}\int d\bm{r}^\prime d\bm{r} \rho(\bm{r}^\prime)V(\bm{r}^\prime-\bm{r})\rho(\bm{r}) = \frac{1}{N_{\mathrm{uc}}}\sum_{\bm{R}^\prime\bm{R}}\int d\bm{r}^\prime d\bm{r}|\phi_{\bm{R}^\prime}(\bm{r}^\prime)|^2 V(\bm{r}^\prime-\bm{r})|\phi_{\bm{R}}(\bm{r})|^2
\end{equation}
contains the interaction of an electron in a Gaussian site with its own charge density (the term $\bm{R}=\bm{R}^\prime$ in the sum), which is unphysical since each site contains only one electron (c.f. the self-interaction energy in density functional theory).
The role of the Fock term this limit is simply to precisely cancel out this self-interaction.
Using this insight, we can compute the Fock term as simply the negative of the self-interaction term.
Since each $\bm{R}$ is equivalent, we can consider only a single $\bm{R}=\bm{0}$, giving the Fock energy per electron as
\begin{equation}
E_F =  -\int d\bm{r}^\prime d\bm{r}|\phi_{\bm{0}}(\bm{r}^\prime)|^2 V(\bm{r}^\prime-\bm{r})|\phi_{\bm{0}}(\bm{r})|^2 
\end{equation}
Let us define the Gaussian charge density $n(\bm{r})=|\phi_{\bm{0}}(\bm{r})|^2=\frac{2\xi^2}{\pi}e^{-2\xi^2|\bm{r}|^2}$ and its Fourier transform ${n}(\bm{k})=\int d\bm{r} n(\bm{r})e^{-i\bm{k}\cdot\bm{r}}=e^{-|\bm{k}|^2/(8\xi^2)}$.
Then, the Fock energy is simply
\begin{equation}
E_F =  -\int \frac{d\bm{k}}{4\pi^2} {n}(\bm{k})^2 V(\bm{k}) 
\end{equation}
where $V(\bm{k})=\int d\bm{r} V(\bm{r})e^{-i\bm{k}\cdot\bm{r}}$ is the Fourier transform of the interaction potential.
In the manuscript, we considered bare Coulomb interaction, which resulted in the effective interaction $V(\bm{k})=\frac{V_C}{|\bm{k}|}e^{-\frac{\pi N}{\Omega_{\mathrm{BZ}}}|\bm{k}|^2}$ for the corresponding zero-$\mathcal{B}$ problem.
With this effective interaction, the Fock energy can be integrated directly to give
\begin{equation}
E_F =  -\int \frac{d\bm{k}}{4\pi^2}  e^{-\frac{|\bm{k}|^2}{4\xi^2}} \frac{V_C}{|\bm{k}|}e^{-\frac{\pi N}{\Omega_{\mathrm{BZ}}}|\bm{k}|^2} = -\frac{V_C\xi^2/(4\sqrt{\pi})}{\sqrt{1+\frac{4 \pi N \xi^2}{\Omega_{\mathrm{BZ}}}}}
\end{equation}
The expression for $E_K + E_F$ with $N=1$ is presented in the main text.

\end{document}